\documentclass[a4paper,11pt]{article}
\usepackage{pos}

\title{The Hubble parameter of the Local Distance Ladder from dynamical dark energy with no free parameters}

\author*[a,b]{Maurice H.P.M. van Putten}

\affiliation[a]{Physics and Astronomy, Sejong University,\\
  Yeongsil-Gwan 614, 209 Neungdong-ro, Seoul, South Korea}

\affiliation[b]{INAF-OAS,\\
via P. Gobetti, 101, I-40129 Bologna, Italy}

\emailAdd{mvp@sejong.ac.kr}

\abstract{
Our cosmology contains Big Bang relic fluctuations by a loss of time-translation symmetry on a Hubble time scale. The contribution to the vacuum is identified with dynamical dark energy $\Lambda\simeq \alpha_p\Lambda_0$ by an IR coupling $\alpha_p\sim \hbar$ of the bare cosmological constant $\Lambda_0\sim\hbar^{-1}$ consistent with general relativity, where $\hbar$ is the Planck constant. 
Described by the trace $J=(1-q)H^2$ of the Schouten tensor derived from a path integral formulation with gauged global phase, 
the proposed $J$CDM takes us beyond the $\Lambda$CDM limit of frozen $J=\Lambda$. 
The Hubble constant $H_0$ in $J$CDM is effectively $\sqrt{6/5}$ times the {\em Planck} value in $\Lambda$CDM analysis of the CMB according to $H(z)=H_0\sqrt{1+(6/5)\Omega_{M,0} Z_5(z) + \Omega_{r,0}Z_6(z)}/(1+z)$, where $Z_n=(1+z)^n-1$ given densities of matter $\Omega_{M,0}$ and radiation $\Omega_{r,0}$. 
With no free parameters, $J$CDM hereby agrees with the Local Distance Ladder when satisfying the BAO measured by {\em Planck}. 
On this cosmological background,
galaxies possess an essentially $C^0$-transition to anomalous dynamics due to reduced inertia below the de Sitter scale of acceleration $a_{dS}=cH$, 
where $c$ is the velocity of light. This is confirmed in SPARC over a 6$\sigma$ tension in $\Lambda$CDM galaxy models, pointing to ultra-light CDM of mass $m_Dc^2<3\times 10^{-21}$eV.
Sensitivity to this cosmological background explains the JWST 'Impossible galaxies' at cosmic dawn by rapid gravitational collapse. 
We comment on an outlook on future confrontations with observations by {\em Euclid}.}

\FullConference{Corfu Summer Institute 2023 "School and Workshops on Elementary Particle Physics and Gravity" (CORFU2023)\\
 23 April - 6 May , and 27 August - 1 October, 2023\\
Corfu, Greece\\}

\begin{document}

\maketitle

\section{Introduction}

The formation and evolution of the large scale structure in galaxies show the central role of gravitation in the Universe, on the Hubble scale down to gravitational collapse to black holes \citep{gen03,ghe03,ghe08,gen10,abb16,EHT22}. The Universe today, however, reveals a dramatically distinct aspect in accelerated expansion \citep{agh20,rie22}. While the former builds on Newtonian gravitational interactions extended in general relativity, accelerated expansion takes us in a different direction with {\em conformal scaling} of cosmological spacetime by a Friedmann scale $a(t)$ in the Friedmann-Lema\^itre-Robertson-Walker (FLRW) line-element
\begin{eqnarray}
    ds^2 = a^2\eta_{ab}dx^a dx^b,
    \label{EQN_01}
\end{eqnarray}
where $\eta_{ab}$ is the metric of Minkowski spacetime. 
With Hubble parameter $H=\dot{a}/a$ in the equivalent line-element $ds^2 = -c^2dt^2+a(t)^2\left(dx^2+dy^2+dz^2\right)$ with velocity of light $c$,
(\ref{EQN_01}) describes a Big Bang cosmology marked by a singularity at a Hubble time $1/H_0$ in the past \citep[e.g.][]{oli22} for a Hubble constant $H_0$ today. 
Conventionally, $a(t)$ in (\ref{EQN_01}) evolves according the Friedmann equations with constant dark energy $\Lambda$ and a finite dimensionless density $\Omega_{M,0}$ of cold baryonic and dark matter - the concordance model  of $\Lambda$CDM. Observations of
the Hubble expansion in late-time cosmology shows $\Lambda>0$ based on a deceleration parameter $q_0<0$ \citep{per99,rie98}, 
where $q(t)=-a\ddot{a}/\dot{a}^2$, equivalently, $q(z)=-1+(1+z)H^{-1}(z)H^\prime(z)$ as a function of cosmological redshift $z$ defined by $a/a_0=1/(1+z)$. 

The concordance model $\Lambda$CDM describes the evolution of $a(t)$ in the presence of a constant dark energy $\Lambda$ and cold dark matter. $\Lambda$ introduces a conformal scaling by $a(t)$ by which conventional null-infinity in $\eta_{ab}$ is lost. 
Such may have consequences for the future state of the Universe. 
If the future state is distinct from what is expected in $\Lambda$CDM, $\Lambda$CDM cannot hold to all orders today, 
inevitably leaving finite tensions between $\Lambda$CDM and the Local Distance Ladder \citep{rie22}. 
A constant $\Lambda$ in $\Lambda$CDM predicts a future de Sitter state with deceleration parameter $q=-1$, though
it represents a singular limit of (\ref{EQN_01}) that might not be stable.  

In a Big Bang quantum cosmology, the Universe caries a background of relic fluctuations by a broken global time-translation invariance on a Hubble scale. Furthermore, the total number of degrees of freedom $N$ within a Hubble volume will be finite. This suggests a background de Sitter energy \citep{van20}
\begin{eqnarray}
Q = N\epsilon
\label{EQN_Q},
\end{eqnarray}
summed over the energy $\epsilon \sim H\hbar$ per degree of freedom by the Heisenberg uncertainty principle.
In a consistent IR limit of quantum cosmology as the Planck constant $\hbar$ approaches zero, 
a finite $Q$ requires a scaling $N\sim 1/\hbar \sim 1/l_p^2$ in Planck units $l_p^2$, $l_p=\sqrt{G\hbar/c^3}$ given
Newton's constant $G$.
In the absence of any other intrinsic length scale, 
the total count $N\sim A_p$, where $A_p=4\pi R_H^2/l_p^2$ denotes the Hubble area in Planck units. 
Conventionally, we consider the Bekenstein bound $N=(1/4)A_p$.

Accordingly, (\ref{EQN_01}) contains a de Sitter heat (\ref{EQN_Q}) in a consistent IR limit of quantum cosmology, 
even as a detailed description thereof remains elusive. 
A discussion on the converse in a consistent UV-completion of the IR limit in modeling (\ref{EQN_01}) has been developed for EFT by the swampland conjectures \citep{obi18,agr18,col19,gar19}. 

By the first law of thermodynamics, $Q$ can be identified with heat. By $dQ=TdS-pdV$, $k_BT=\lambda_1 H\hbar$ per degree of freedom and $p=-\rho_c$, $dE = \left| p \right| dV - TdS$, $S=k_BN$ given the Boltzmann constant $k_B=1.38\times 10^{-16}$eerg\,K$^{-1}$.
The energy $\epsilon = k_BT_H$ in each surface element of ${\cal H}$ derives from the horizon temperate $k_BT_H = \lambda_1 \hbar c/R_H$. It satisfies
$dE = \rho_c A_H dR_H - \epsilon dN = c^{-1}L_0 \left[ \frac{3}{2} - 2\pi \lambda_1 \right] dR_H$, where $L_0=c^5/G$.
Energy conservation $dE=dMc^2$ fixes $\lambda_1 = 1/2\pi$, 
where $R_H=2R_g$ given the gravitational radius $R_g=GM/c^2$ of the total mass-energy within $R_H$. 
Hence, $\epsilon =   \hbar c/2\pi R_H$ at the de Sitter temperature $k_BT_{dS}=H\hbar/2\pi$ - recovering \citep{gib77}.  
See further \citep{pad05} on the role of the first law of thermodynamics in Friedmann cosmologies and \S3.1 below.  

Averaging $Q=NH\hbar/2\pi$ over a Hubble volume $V_H=(4\pi/3)R_H^3$ recovers the closure energy density
\begin{eqnarray}
\rho_c = \frac{Q}{V_H} = \frac{3c^4}{8\pi G R_H^2},
\label{EQN_rhoc}
\end{eqnarray}
where $R_H=c/H$ denotes the Hubble radius. $\rho_c$ is hereby equivalent to the energy density $Q/V_H$ in heat $Q$ of a black hole of Schwarzschild radius $R_S$, averaged over the Schwarzschild volume $V_H=(4\pi/3)R_S^3$, based on the Clausius integral $Q=\int_0^M T_H dS=Mc^2$ of the Hawking temperature $T_H$ over the Bekenstein-Hawking entropy $S=k_BN$ of ${\cal H}$.

In the above, general relativity is considered to be the limit of a theory of quantum cosmology in the limit as $\hbar$ approaches zero. 
The elliptic part is hereby left scale-free, while the hyperbolic part possesses $L_0=c^5/G=m_pc^2/t_p$, where $m_pc^2=\hbar c/l_p$ and $t_p=l_p/c$ refer to the Planck mass and, respectively, Planck time. The first recovers the classical limit of black holes given the luminosity of evaporation $L_H\sim \hbar$, while the second defines a scale for the peak luminosity $\hat{L}_{GW}$ in gravitational radiation from binary black hole coalescence, notably demonstrated by $\hat{L}_{GW}\simeq 0.1L_0$ in GW150914 \citep{abb16a}.

Applied to quantum fluctuations of the vacuum, {\em consistent IR coupling} in general relativity requires 
normalization of the UV-divergent Zeldovich bare cosmological constant $\Lambda_0=8\pi Gc^{-4}\rho_0 \sim 1/\hbar$ 
from a Planck density $\rho_0=\hbar c/l_p^4$. 
Such quantities are {\em primitives}, whose consistent IR coupling in general relativity call for a dimensionless coupling 
over $\sim \hbar$. Since $l_p^2\sim \hbar$, this coupling equivalently obtains by normalization to surface area in Planck units.
With no intrinsic scale of area in general relativity, such normalization 
in (\ref{EQN_rhoc}) derives from 
the IR coupling relation \cite{van24d}
\begin{eqnarray} 
\alpha_p A_p=1,
\label{EQN_IR}
\end{eqnarray}
whereby
\begin{eqnarray}
\Lambda \equiv \alpha_p\Lambda_0 = 2H^2/c^2.
\label{EQN_L0}
\end{eqnarray}

Given the dimensionless matter density $\Omega_{M,0}\simeq 0.3$ at the present epoch \citep{agh20}, indeed the present epoch calls for a vacuum contribution (\ref{EQN_L0}) in a three-flat cosmology.
By the first Friedmann equation, a future de Sitter state and, by implication, $\Lambda$CDM is non-existent in the distant future. In fact, $H(z)$ may be divergent consistent with a Hubble constant $H_0$ in the Local Distance Ladder being greater than the {\em Planck} value, e.g., when de Sitter is unstable \citep{abc21}.

To study the observational consequences of (\ref{EQN_01}) in the IR limit (\ref{EQN_IR}) of a Big Bang quantum cosmology, 
dark energy (\ref{EQN_Q}) and (\ref{EQN_L0}) is identified with the trace of the Schouten tensor \citep{sch11}, 
seen to derive from gauging global phase in a path integral formulation of quantum cosmology (\S2). 
An equivalent interpretation is found in some thermodynamic properties of spacetime (\S3).
The resulting Hubble expansion $H(z)$ ameliorates multiple tensions in observations of late-time cosmology (\S4), effectively related by scaling of {\em Planck} $\Lambda$CDM values.

Next, we turn to the JWST `Impossible galaxies', upending $\Lambda$CDM at ultra-high redshft. 
This is unlikely solved by cosmic variations of fundamental constants (\S5). 
Rather, it may derive from anomalous inertia in the limit of weak gravitation at accelerations $\alpha$ below the de Sitter scale of acceleration
\begin{eqnarray}
a_{dS}=cH
\label{EQN_adS}
\end{eqnarray}
of the cosmological background vacuum. 
To this end, curvature in general relativity is identified with consistent IR coupling (\ref{EQN_IR}) to position information of particles based on their propagator (\S6) - a second primitive in the sense of UV-divergence in $\hbar$. 
On a cosmological background (\ref{EQN_adS}), 
weak gravitation at accelerations $\alpha < a_{dS}$ is seen to yield early galaxy formation by fast gravitational collapse on the same footing as the baryonic Tully-Fisher relation (\S7). We summarize our findings in \S8.

\section{Beyond dark in $\Lambda$CDM}

Einstein's theory of general relativity contains a scalar theory of Newton'd theory of gravitation by coupling matter to the scalar curvature $R$ of the Ricci tensor $R_{ab}$. 
Specifically, it couples $R$ to the trace $\rho=-T$ of a matter stress-tensor $T_{ab}$ in
\begin{eqnarray}
R=-\kappa T,
\label{EQN_R}
\end{eqnarray}
where $\kappa=8\pi$ in geometrical units $G=c=1$. 
The vacuum equations $R=0$ famously mixed elliptic-hyperbolic, giving rise to black holes and gravitational-wave propagation. 
Einstein equations coupling $R_{ab}$ to the full tensor $T_{ab}$, i.e.,
$R_{ab}+\lambda_0 g_{ab}R=\kappa T_{ab}$ preserves (\ref{EQN_R}), provided that $1+4\lambda_0=-1$, i.e., $\lambda_0 = -\frac{1}{2}$. 
Newton's theory in (\ref{EQN_R}) is hereby embedded in 
\begin{eqnarray}
 G_{ab}=\kappa T_{ab},
\label{EQN_G0}
\end{eqnarray}
where $G_{ab}=R_{ab}-(1/2)g_{ab}R$ is the Einstein tensor. 
$G_{ab}$ is divergence free, whereby (\ref{EQN_G0}) conserves energy-momentum in $\nabla^aT_{ab}=0$. 
By the embedding of (\ref{EQN_R}) in (\ref{EQN_G0}), the latter equivalently satisfies
\begin{eqnarray}
G_{ab} + \mu g_{ab}R = \kappa \left(T_{ab} -\mu g_{ab} T\right),
\label{EQN_G1}
\end{eqnarray}
where $\mu$ is abritrary.

Einstein equations (\ref{EQN_G0}-\ref{EQN_G1}) conventionally refer to baryonic content in matter. 
The content of the Universe, however, appears predominantly dark with no observable interactions with baryons other than by gravitation. What, then, is a natural coupling between the dark sector and spacetime?

Consider a stress-energy tensor $M_{ab}$ of the dark sector of the Universe with essentially no interactions with the standard model particles, $\nabla^aM_{ab}=0$.
Einstein's cosmological term $M_{ab}=\Lambda g_{ab}$ is a special case with $\Lambda$ constant. $M_{ab}$ can be expanded into matter (on-shell) and a trace $K$,
\begin{eqnarray}
{M}_{ab}=T_{ab}- Kg_{ab}.
\label{EQN_M0}
\end{eqnarray}
Equivalently, $R_{ab} = \kappa \left( {M}_{ab} - \frac{1}{2}g_{ab}{M}\right)$ by $\mu=1/2$ in (\ref{EQN_G1}), whereby
\begin{eqnarray}
   R_{ab} =  \kappa \left( {T}_{ab} - \frac{1}{2}g_{ab}{T}\right) 
    + \kappa K g_{ab}.
    \label{EQN_M1}
\end{eqnarray}
Schouten considers an expansion $R_{ab} = 2P_{ab} + J g_{ab}$ in $P_{ab}$ and its trace $J$ \citep{sch11}.
In four-dimensional spacetime, $J$ satisfies
\begin{eqnarray}
    J=\frac{1}{6}R.
    \label{EQN_J1}
\end{eqnarray} 
By (\ref{EQN_M1}) and (\ref{EQN_J1}), we identify
\begin{eqnarray} 
P_{ab} = \frac{1}{2}\kappa \left( T_{ab} - \frac{1}{2} g_{ab}T\right),~\kappa K=J. 
\label{EQN_J2}
\end{eqnarray}
Tracing back, Einstein equations for the dark sector satisfies 
\begin{eqnarray}
   G_{ab} = \kappa  T_{ab} - J g_{ab}.
\label{EQN_J3}
\end{eqnarray}
The right hand-side of (\ref{EQN_J3}) may further be identified with 
$T_{ab}=\left[(1-q)\pi_{ab}^++q\pi_{ab}^-\right]\rho_c$, where 
$\pm\pi_{ab}^\pm={\rm dia}\left[1,\pm1/3,\pm1/3,\pm1/3\right]$,
${\rm tr}\,\pi_{ab}^\pm = 1 \mp\left(-1\right)$ in the metric
signature $\left(-,+,+,+\right)$ \citep{van20}.

\subsection{Dark energy from gauging total phase}

Our formulation (\ref{EQN_J3}) identifies a dynamical dark energy
\begin{eqnarray}
    \Lambda = J
    \label{EQN_L1}
\end{eqnarray}
in (\ref{EQN_J1}) and (\ref{EQN_J4}).
Following \S1, (\ref{EQN_L1}) originates as relic fluctuations of the Big Bang, breaking time-translation symmetry on a Hubble scale. 

A further derivation follows from the path integral formulation, following a foliation of spacetime in Cauchy surfaces of constant cosmic time $t$. The result has an interpretation in de Sitter heat (\ref{EQN_Q}),
consistent with aforementioned first law of thermodynamics.

General relativity satisfies the action principle applied to the Hilbert action $S=\int L\sqrt{-g}d^4x$ for a Lagrangian $L=R+\cdots$, including additional terms from the standard model. 
The Einstein equations obtain as classical equations of motion from the variational principle {\em with respect to sub-Horizon scale fluctuations}. 
For super-horizon fluctuations, the action principle calls for a global phase reference $\Phi_0$ in the propagator - the gauging of a global symmetry \citep{led09,van20,cha20,ber15,sad22}. 

These super-horizon scale fluctuations are off-shell, not localized in the visible Universe within the Hubble horizon ${\cal H}$. New contributions derive from the propagator gauged by total phase \citep{van20},  
\begin{eqnarray}
    e^{i\Phi} \rightarrow e^{i\left(\Phi - \Phi_0\right)},
    \label{EQN_L2}
\end{eqnarray}
where $\Phi = S/\hbar$. 

Conventionally, the total phase $\Phi_0 = \Phi_0\left[{\cal N}\right]$ is safely put to zero, assuming asymptotical null-infinity ${\cal N}$ of Minkowski spacetime. Correspondingly, the conformal scale $a=a_0$ in (\ref{EQN_01}) is {\em frozen} and the classical Einstein equations obtain from the standard (scale-free) variational principle. In (\ref{EQN_01}), however, $a=a(t)$ is time-dependent, and the same boundary condition does not apply. In the face of 
${\cal H}$, the variational principle is no longer scale-free and super-horizon scale fluctuations exist call for a gauge 
\begin{eqnarray}
 \Phi_0 = \Phi_0\left[ {\cal H}\right].
 \label{EQN_L3a}
\end{eqnarray}
This gauge is dynamical since $R_H$ is time-dependent. 
It can be absorbed in above-mentioned Lagrangian $L$ by an equivalent density $2\Lambda$ with
\begin{eqnarray}
 \Lambda = \lambda_2 R = g J
 \label{EQN_L3b}
\end{eqnarray}
for some constant $g=6\lambda_2$.
A detailed confrontation with data preserving the astronomical age of the Universe, consistent with $\Lambda$CDM, and the BAO, shows \citep{van21}
\begin{eqnarray}
    g\simeq 0.9964.
    \label{EQN_L3c}
\end{eqnarray}
{\em Data identify dark energy (\ref{EQN_L3b}) to be dynamical and non-local, derived from gauging global phase in the propagator, described by the trace $J$ of the Schouten tensor.}

\subsection{Dark energy: local or non-local?}

In the conformally flat Friedmann cosmology (\ref{EQN_01}), 
(\ref{EQN_J1}) reduces to $J=(1-q)H^2$, i.e., 
\begin{eqnarray}
    J = \frac{1}{2}a^{-2}\frac{d^2}{dt^2} a^2.
    \label{EQN_J4}
\end{eqnarray}
A constant $J$ would recover a de Sitter solution $a(t)=a_0 e^{H_0t}$, e.g., $H_0 = \sqrt{\Lambda/3}$ assuming $R=4\Lambda$, but such is ruled out by (\ref{EQN_L0}).
By (\ref{EQN_J1}), (\ref{EQN_J3}) suggests an equivalent formulation in
\begin{eqnarray}
    H_{ab} \equiv R_{ab} - \frac{1}{3}g_{ab} R = \kappa T_{ab}. 
    \label{EQN_H0}
\end{eqnarray}
In (\ref{EQN_H0}), $H_{ab}$ nor $T_{ab}$ is divergence free, only the total (\ref{EQN_M0}) is divergence free. 
A baryonic component $T_{ab}$ can be considered alongside dark matter $M_{ab}$ in the right hand-side of (\ref{EQN_H0}) by the substitution $T_{ab} \rightarrow  T_{ab} + M_{ab}$.
$T_{ab}$ remains divergence free since $T_{ab}$ and $M_{ab}$ are non-interacting, except by gravitation. 
With a trace-adjusted coupling to curvature for baryons, (\ref{EQN_H0}) becomes
\begin{eqnarray}
    H_{ab} = \kappa \left(T_{ab}-\frac{1}{6}g_{ab} T\right) + 
    \kappa M_{ab}.
    \label{EQN_H1}
\end{eqnarray}
By (\ref{EQN_G1}) with $\mu=1/6$, (\ref{EQN_H1}) preserves the Einstein equations whenever matter density 
substantially exceeds closure density.
On average and away from these regions, $M_{ab}$ is dominant given the 
relatively small cosmological content $\Omega_b\simeq 5\%$ of baryons overall. The trace adjustment in (\ref{EQN_H1}) hereby differs from dark matter alone in (\ref{EQN_H0}) by less than 1\%.
The formulation (\ref{EQN_H0}-\ref{EQN_H1}) derived from (\ref{EQN_J1}) goes beyond $\Lambda$CDM - the limit $J\equiv \Lambda$ frozen to a constant ignoring (\ref{EQN_Q}). 

However, (\ref{EQN_H1}) implicitly allows $J=R/6$ to be a local scalar in $H_{ab}$, distinct from 
$J=(1-q)H^2/c^2$ in (\ref{EQN_01}) 
representing the non-local properties of the cosmological vacuum inferred from gauging global phase (\ref{EQN_L3a}).
While classical sources satisfy causality, contributions from fluctuations on super-horizon scales point the other way. 
To be sure, $\Lambda=J$ is non-local in the Hilbert action, which appears inevitable in the IR limit of quantum gravity (\ref{EQN_IR}). 
Perhaps this is not surprising, since (\ref{EQN_Q}) originates in the breaking of an essentially global symmetry 
on a Hubble time-scale inherent to a Big Bang quantum cosmology. When modeling structures on sub-horizon scales,
these arguments favor (\ref{EQN_J3}) with mixed local and non-local contributions explicitly in $T_{ab}$ and, 
respectively, $J$, rather than (\ref{EQN_H0}).

\section{Heat and observed temperature}

The dark energy density (\ref{EQN_L1}) in the previous section identifies a background distribution of 
de Sitter-Schouten heat. It identifies a finite temperature of cosmological spacetime \citep{gib77} in a background of fluctuations by a dynamic dark energy (\ref{EQN_Q}-\ref{EQN_rhoc}).

\subsection{De Sitter-Schouten heat in cosmological spacetime}

We identify (\ref{EQN_L1}) with the first law of thermodynamics and the Unruh temperature 
$k_BT_H=a_H\hbar/2\pi c$ inferred from surface gravity $a_H$ \citep{van15b} as follows.
Consistent with the first law of thermodynamics (\S1), 
the de Sitter temperature $T_{dS}$ is the thermodynamic temperature
\begin{eqnarray}
    T_{dS}=\left(\frac{\partial S}{\partial E}\right)^{-1} = \frac{H\hbar}{2\pi c}.
    \label{EQN_TdS}
\end{eqnarray}
Here, $S=k_BI$ for the position information $I=2\pi\varphi$ of a particle of mass-energy $E$ with
Compton phase $\varphi = ER_H/\hbar c$ relative to the Hubble horizon ${\cal H}$ at Hubble radius $r=R_H$, elucidated further in \S6 below. 
The Bekenstein bound \cite{bek81} applied to a 2-sphere of radius $r<R_H$ corresponds to 
$I\le \frac{1}{4}A_p$, which recovers the familiar limit $r\ge 2R_g$, $R_g=GE/c^4$.
Following \S1, variations in the Hubble radius carry an associated change of heat $dQ=T_HdS_H$, 
$S_H=\frac{1}{4}k_BA_p$ of ${\cal H}$,
\begin{eqnarray}
    dQ = \frac{1}{4l_p^2}\left(\frac{a_H\hbar}{2\pi c}\right)\left(8\pi R_H dR_H\right) = \frac{\hbar a_H}{cl_p^2}R_HdR_H.
\end{eqnarray}
Attributed to work by pressure $p=-\rho_\Lambda$ 
from a dark energy density $\Lambda = \left(8\pi G/c^4\right)\rho_\Lambda$, we have
\cite[cf.][]{eas11}
\begin{eqnarray}
    \Lambda  = \frac{8\pi G}{c^4} \frac{dQ}{4\pi R_H^2 dR_H} = 2c^{-2}\frac{a_H}{R_H}.
    \label{EQN_la}
\end{eqnarray}
By $\Lambda=J$ and $J=(1-q)H^2/c^2$ of \S2.1-2, we infer
\begin{eqnarray}
    a_H = \frac{1}{2}\left(1-q\right)a_{dS}
    \label{EQN_aH}
\end{eqnarray}
by (\ref{EQN_adS}). The result identifies the temperature
\begin{eqnarray}
T_H = \left(\frac{1-q}{2}\right) T_{dS}
\label{EQN_TH}
\end{eqnarray}
of an evolving horizon of a Big Bang cosmology (\ref{EQN_01}). 

\begin{figure*}
    \centering
    \includegraphics[scale=0.55]{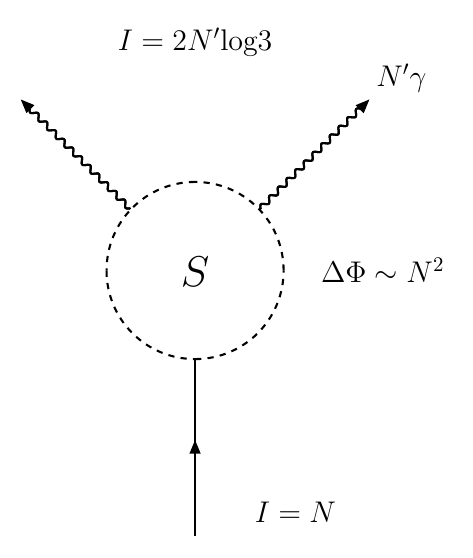}
    \hskip0.5in\includegraphics[scale=0.18]{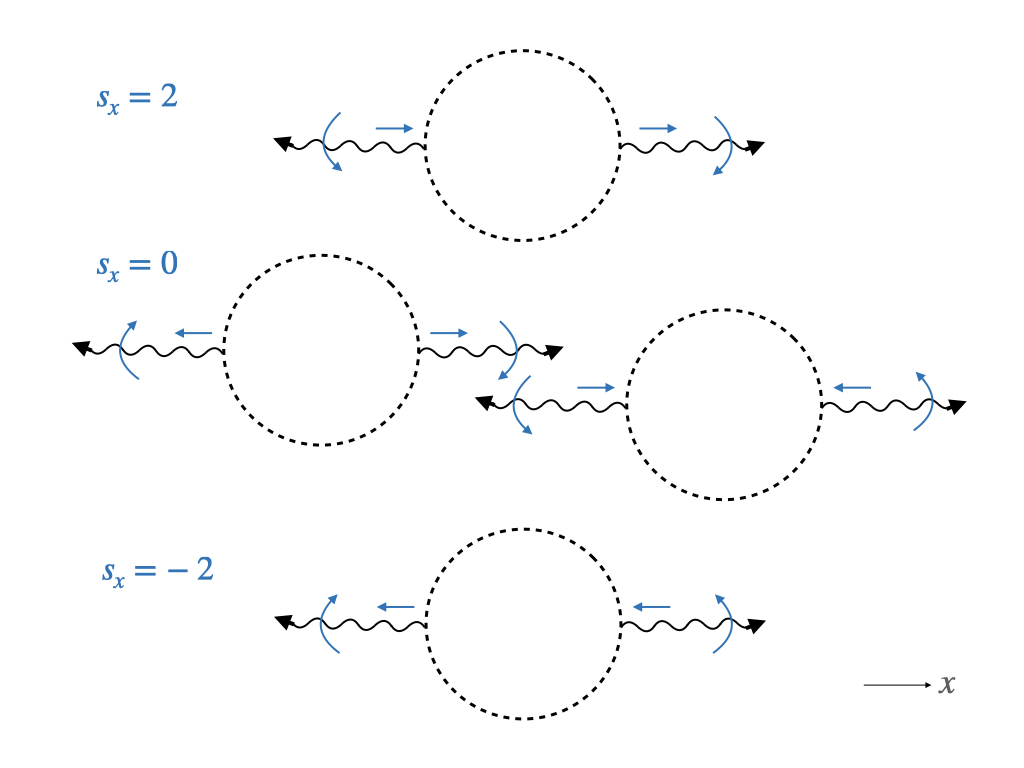}
    \caption{
    (Left.) A black hole of mass $M$ can be viewed as a finite temperature quantum computer for a unitary transformation - hidden from the observer - of information $I=k_BN$ in $N$ Planck sized units of area over a change in total phase $\Delta \Phi \sim N^2$. It satisfies $\Delta \Phi \sim M^2R^2$ at the Margolus-Levitin bound $\Delta \Phi \sim Mt_{ev}$ \citep{mar98} radiating $I$
    in an equivalent number $N^\prime$ of 2PE events over its lifetime $t_{ev}\sim M/L$ at a luminosity $L$, $Lt_c^2\sim \hbar$, given the light crossing time $t_c=R/c$ for a radius $R$. This identifies $N\sim MR$. For a black body, this leaves $M\sim NT$ at temperature $T\sim 1/R$, whence $N\sim R^2$ and $R\sim M$. Distinct from a baryonic object, the horizon releases $i=2k_B\log3$ per 2PE at the observed effective temperature of one-half $M$'s temperature in antipodal pairs $\left(\gamma,\gamma\right)$ by $\mathbb{P}^2$ topology.
    (Right.) A $\mathbb{P}^2$ invariant reading of information in entangled 2PE obtains from $\log3$ in the principle directions and $\log3$ in the sum of spins in each of the photons according to their helicity. Schematically shown are the three possible sums $s={-2,0,2}$ for propagation along the $x$-axis.
    }
    \label{FIGa}
\end{figure*}

\subsection{Topology and temperature of event horizons} 

The (modified) de Sitter temperature (\ref{EQN_TH}) of heat (\ref{EQN_Q}) in spacetime refers to virtual fluctuations of spacetime background, distinct from (on-shell) Hawking radiation. 
This distinction originates in a temperature ambiguity in Hawking radiation \citep{tho84,tho85,tho21}. Since Hawking radiation is thermal, it is conventionally interpreted as radiation from a black body with equal temperature of radiation and the emitting surface. 
While the event horizon of a black hole is classically black, {\em without baryonic matter such temperature equality need not hold identically} (Fig. \ref{FIGa}).

By $\mathbb{P}^2$ of black hole horizons, we have \citep{van24c}
\begin{eqnarray}
T_{BH}=2T_\gamma 
\label{EQN_TG1}
\end{eqnarray}
between the temperature $T_{BH}$ of the black hole spacetime
and $T_\gamma$ in electromagnetic radiation, here in entangled two-photon emission (2PE) seen by a distant observer \citep{van24c}. 

Here, 2PE is in antipodal by the $\mathbb{P}^2$ topology of the event horizon (Fig. 1), distinct from radiation in individual Hawking particles by one-half the temperature of the underlying energy reservoir in the mass-energy of the black hole. The result projects a quantum of information \citep{van24b}
\begin{eqnarray}
    i=2\log3
\end{eqnarray}
onto the celestial sphere in each entangled 2PE emission event, where one $\log3$ is accounted for by orientation in
space and another, independently, by total photon spin in these pairs. The appearance of $\log3$ rather than $\log2$ is 
consistent with detailed considerations of quasi-normal modes and the uniform area spectrum of black holes \citep{hod98,dre03}.

By $S^2$ topology, instead of (\ref{EQN_TG1}), 
the cosmological horizon temperature satisfies equality in
\begin{eqnarray}
    T_{dS}=T_{G},
    \label{EQN_TG2}
\end{eqnarray}
where $T_G$ is the temperature of spacetime.
By (\ref{EQN_TG1}-\ref{EQN_TG2}), we  recover the universal relation
\begin{eqnarray}
   T_G S=Mc^2
   \label{EQN_TG3}
\end{eqnarray}
for a Hubble ($T_G=T_{dS}$) and Schwarzschild horizons ($T_G=T_{BH}$), enclosing a mass $M$ with entropy $S$ according to their surface area. 
By distinct topology, $T_G$ is hereby seen to resolve the well-known temperature discrepancy by a factor of two between the Hawking temperature of a black hole and the temperature of de Sitter space. 
Accordingly, the de Sitter temperature (\ref{EQN_TH}) of (\ref{EQN_Q}) is the temperature of the background spacetime, as does $T_G$ for a black hole.

The spacetime temperature $T_G$ in (\ref{EQN_TG3}) satisfies the holographic scaling \citep[e.g.][]{abc21}
\begin{eqnarray}
    \rho_H = \frac{Mc^2}{\frac{4\pi}{3}R_S^3} = \frac{3c^4}{8\pi R_S^2}\sim T_G^2,
    \label{EQN_TG4}
\end{eqnarray}
satisfying (\ref{EQN_rhoc}) and showing a scaling with temperature squared,
distinct from on-shell radiation energy density scaling with the fourth
power of temperature. 

\section{Hubble expansion $H(z)$}

By the relatively small trace adjustment in the contribution of baryons in (\ref{EQN_H1}), we next consider (\ref{EQN_H0}) with $T_{ab} =\left(\rho_M+p\right)u_au_b+pg_{ab}$ of matter density $\rho_M$ and phantom pressure $p$.
$\rho_M$ is strongly coupled to $J$, and only their combination is divergence free.
In what follows $\Omega_M=\rho_M/\rho_c$ denotes the total dimensionless matter density. 

While $\Lambda$CDM posits the Hamiltonian energy constraint expressed by the first Friedmann equation - the Hamiltonian energy constraint - to be first order in $\dot{a}$, $J$ in (\ref{EQN_H0}) changes its character to second order in time, explicitly shown in (\ref{EQN_J4}). 
A detailed calculation obtains the Hubble expansion rate $H(z)=H_0h(z)$ \citep{van21},
\begin{eqnarray} 
  h_J(z)=\frac{\sqrt{1+\frac{6}{5}\Omega_{M,0}Z_5(z)+\Omega_{r,0}Z_6(z)}}{1+z}
\label{EQN_HJ}
\end{eqnarray} 
with $Z_n(z)=\left(1+z\right)^n-1$. 
In what follows, we shall refer to (\ref{EQN_HJ}) as $J$CDM.

By the $1+z$ denumerator in (\ref{EQN_HJ}), the future de Sitter solution in $z\rightarrow-1$ is unstable, in departure of $\Lambda$CDM. 
$H_0$ in $J$CDM is hereby larger than $H_0$ in $\Lambda$CDM,
evident in $H(z)$ over the future $-1<z<0$ in (\ref{EQN_HJ}).

\subsection{T-duality in $J$CDM}

The divergence in the Hubble expansion $H(z)$ in (\ref{EQN_HJ}) can be anticipated by the T-duality symmetry in the equations of motion.

To this end, consider $\kappa = a_0/a=(1+z)$ in terms of the cosmological redshift $z$. 
($\kappa$ hereby scales with a Rindler acceleration $\xi = c^2/a$.) 
The Friedmann equations can be seen to reduce to \citep{van21}
\begin{eqnarray}
D(\kappa) = 3 \Omega_M, D(a) = - 3\Omega_p
\label{EQN_D}
\end{eqnarray}
in terms of the second order operator $D(u)\equiv \ddot{u}u/\dot{u}^2$, where $p$ refers to total (phantom) pressure.
The solution (\ref{EQN_HJ}) to (\ref{EQN_D}) satisfies exact $T$-duality in the Friedmann scale $a$,
\begin{eqnarray}
    D(a)+D(\kappa) = 2.
    \label{EQN_H3}
\end{eqnarray}
This new symmetry arising from breaking of aforementioned global time-translation symmetry in a Big Bang cosmology.

\subsection{$J$CDM versus $\Lambda$CDM}

In the same notation as (\ref{EQN_HJ}), 
$\Lambda$CDM satisfies 
\begin{eqnarray}
h_\Lambda=\sqrt{1+\Omega_{M,0}Z_3(z)+\Omega_{r,0}Z_4(z)}.
\label{EQN_HL}
\end{eqnarray}
`Comparing, $h_J$ in (\ref{EQN_HJ}) with $h_\Lambda$ in (\ref{EQN_HL}) shows the distinct presence of the factor $6/5$ in matter coupling to background cosmology. 
Preserving the ratio of radiation to matter density at high redshift,
notably the BAO, indicates scaling by 5/6 of matter density compared
to $\Lambda$CDM. Preserving $\Omega_{M,0}h^2$, in turn, indicates an
accompanying scaling by 6/5 of the Hubble parameter in $\Lambda$CDM 
(Table 1), i.e.:
\begin{eqnarray}
\left(H_0,\Omega_{M,0}\right)_J\simeq \left(\sqrt{\frac{6}{5}}H_0,\frac{5}{6}\Omega_{M,0}\right)_\Lambda.
\label{EQN_scaling}
\end{eqnarray}
This anticipates a Hubble tension
\begin{eqnarray}
    \frac{\Delta H_0}{H_0}\simeq \sqrt{\frac{6}{5}}-1\simeq 9.5\%.
    \label{EQN_dH}
\end{eqnarray}
The scaling (\ref{EQN_scaling}) is qualitatively follows the anti-correlation 
between Hubble parameter and matter density according to the degeneracy $\Omega_{M,0}h^3=\,$const in the three-flat cosmologies (\ref{EQN_01}) \citep{oli22}. Based on the {\em Planck} $\Lambda$CDM value $\Omega_{M,0}\simeq 0.31$, (\ref{EQN_dH}), such points to $(5/6)^{3/2}\Omega_{M,0}\simeq0.24$, similar but somewhat lower than $(5/6)\Omega_{M,0}\simeq 0.26$ according to (\ref{EQN_scaling}).

The scaling (\ref{EQN_scaling}) originates in aforementioned non-existence of de Sitter in the face of a finite matter density, as the following shows.

\subsection{BAO analysis of $J$CDM}

The angular scale $\theta_*$ of the BAO in the power spectrum of {\em Planck} CMB
is an essentially model-independent observable \citep{agh20}
 \begin{eqnarray}
     \theta_* = \left(1.04091\pm 0.0003\right)\times 10^{-2}.
     \label{EQN_ts}
 \end{eqnarray}
 Evaluated at recombination redshift $z_*\simeq 1100$, it can be expressed 
 on a given cosmological background by \citep{jed21}
 \begin{eqnarray}
     \theta_* = \frac{\int_{z_*}^\infty \beta_s dz/h(z)}{\int_0^{z_*}dz/h(z)}
      = \beta_{0}\frac{\int_{z_*}^\infty dz/h(z)}{\int_0^{z_*}dz/h(z)},
     \label{EQN_tsI}
 \end{eqnarray}
where $\beta_s=c_s/c\simeq 1/\sqrt{3}$ is the sound speed $c_s$ in the ultra-relativistic baryon-poor epoch prior to the surface of last scattering. Crucially, $\beta_s$ is expected to be essentially invariant for different choices of cosmological background such as $\Lambda$CDM and $J$CDM, when both satisfy the BAO at recombination $z_*$. We are therefore at liberty to use an effective sound speed $\beta_0\simeq 0.80/\sqrt{3}=0.46$ in the ratio on the right hand-side of (\ref{EQN_tsI}), extending a previous analysis on the same \citep{van21}. Indeed, explicit calculations show $\beta_0$ changes by less than 0.5\% in $\Lambda$CDM in varying $H_0$ and $\Omega_{M,0}$ across their respective {\em Planck} uncertainties.

\begin{table}
{\bf Table 1.} Estimates of $\left(H_0,q_0,\Omega_{M,0}\right)$ in $J$CDM fits to to the LDL \citep{van17b} and, independently, to the {\em Planck} BAO angular scale $\theta_*$ (\ref{EQN_ts}), alongside {\em Planck} $\Lambda$CDM values \citep{agh20}. 
$H_0$ is in units of ${\rm km~s}^{-1}{\rm Mpc}^{-1}$. (Reprinted from \cite{van24d}.)
\begin{eqnarray*}
\begin{array}{c|cccccc}
    \hline
    & J{\rm CDM}/{\rm LDL}^a & J{\rm CDM}/{\rm BAO}^b  &  {\rm\sc LDL}^c & \Lambda{\rm CDM}/{\rm CMB}^d & \Lambda{\rm CDM\,scaled}^e\\
    \hline 
    H_0   & 74.9\pm2.6       &    74.7\pm0.91    &  73.04\pm1.04   & 67.36 \pm 0.54 & 73.79\pm 0.59 \\ 
    q_0   & -1.18\pm0.084 &   -1.23\pm0.027 &  -1.08\pm0.29   &-0.5273 \pm 0.011 & -1.21\pm 0.018 \\
    \Omega_{M,0} & 0.2719\pm0.028 & 0.2565\pm0.0063  &  -  & 0.3153 \pm 0.0073 & 0.2628\pm 0.0061\\ 
    \hline
\end{array}
\label{EQN_T1}
\end{eqnarray*}
$a$. Local Distance Ladder of late-time cosmology \citep{van17b};
$b$. Fig. \ref{fig_BAO};
$c$. LDL: $H_0$ \citep{rie22} and $q_0$ \citep{cam20}; 
$d$. Recommended values of {\em Planck} \citep{agh20,lah22};
$e$. Based on $\sqrt{6/5}H_{0}$ and $(5/6)\Omega_{M,0}$ scaling (\ref{EQN_scaling}) of $\Lambda$CDM. 
\end{table}

\begin{figure*}
    \centerline{
    \includegraphics[scale=0.2]{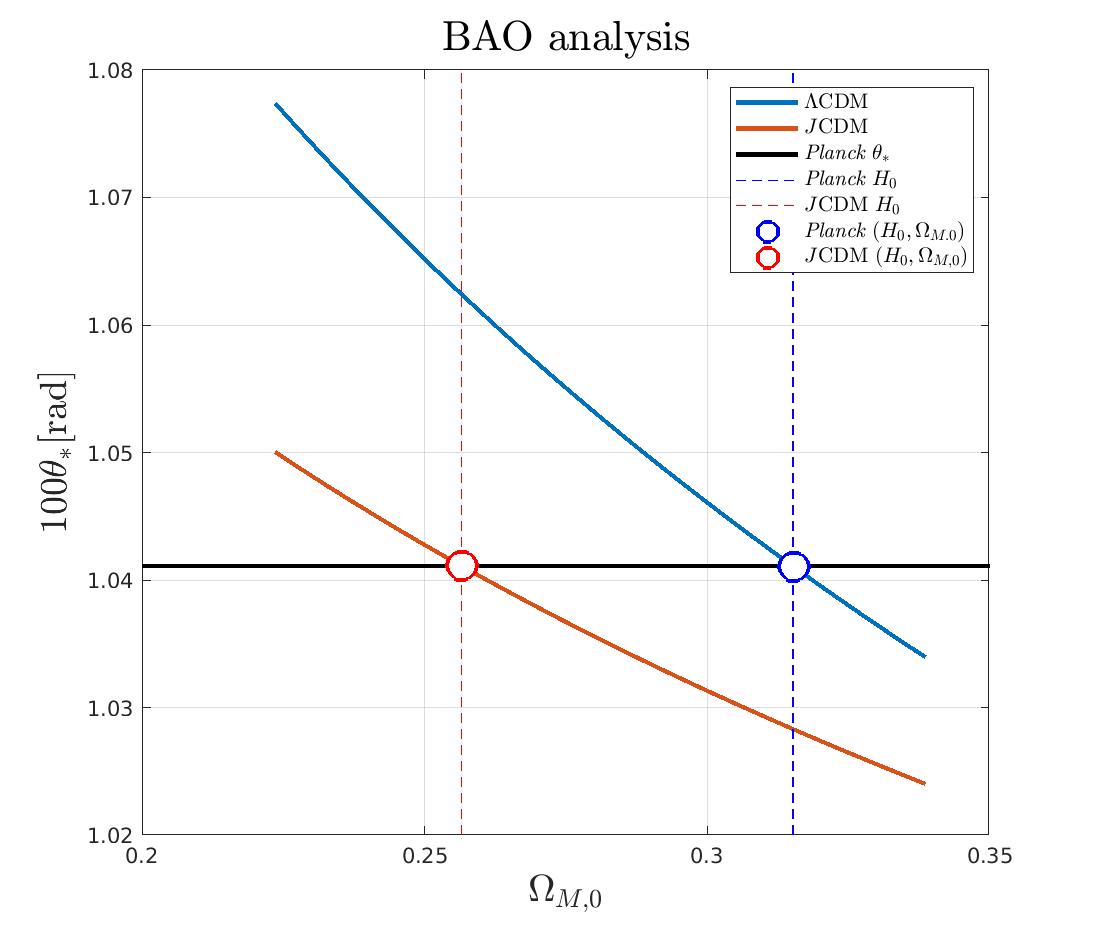}
    \includegraphics[scale=0.2]{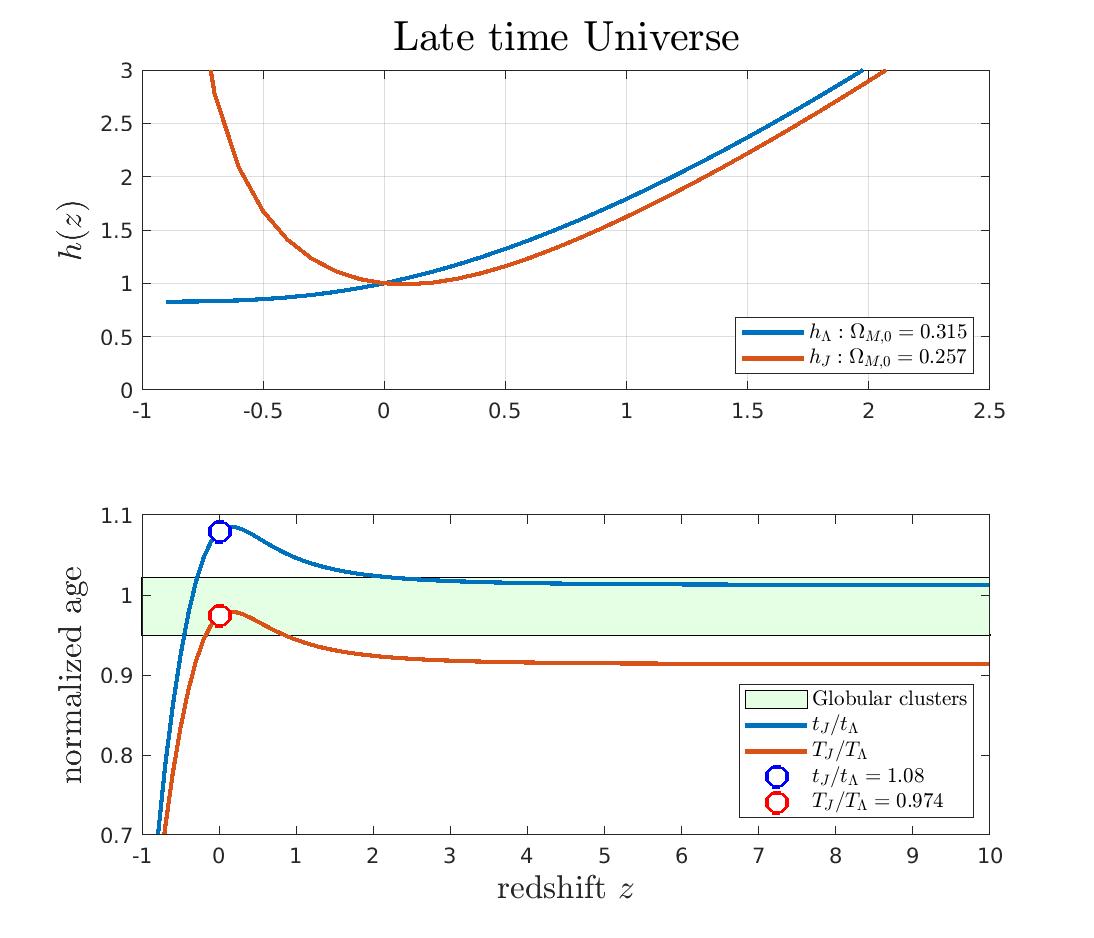}}
    \vskip-0.0in
    \caption{
    (Left Panel.) 
    Cosmological parameters $\left(H_0,\Omega_{M,0}\right)$ for a model background are tightly constrained by the {\em Planck} BAO angle $\theta_*$ (\ref{EQN_ts}) and
    radiation-to-matter ratio (\ref{EQN_MR}).
    $J$CDM satisfies these constraints for $\left(h,\Omega_{M,0}\right)_J=\left(0.747,0.2565\right)$ consistent with the anticipated scaling (\ref{EQN_scaling}) of {\em Planck} values.
    (Right Panels.) $J$CDM preserves the present astronomical age of the Universe according to the oldest stars in globular clusters (green) and $\Lambda$CDM analysis of the CMB. (Reprinted from \cite{van24d}.}
    \label{fig_BAO}
\end{figure*}

We proceed with two constraints: the BAO (\ref{EQN_ts}) by (\ref{EQN_tsI}) and the {\em Planck} ratio of radiation-to-matter density
\begin{eqnarray}
\frac{\Omega_{r,0}}{\Omega_{M,0}}=\frac{\Omega_{r,0}h^2}{\Omega_{M,0}h^2}=\left(1.7238\pm 0.03\right)\times 10^{-4}
\label{EQN_MR}
\end{eqnarray}
of the {\em Planck} values $\Omega_{r,0}h^2 = 2.4661\times10^{-5}$ and $\Omega_{M,0}h^2=0.1431$, 
where $h=H_0/\left(100\,{\rm km}\,{\rm s}^{-1}{\rm Mpc}\right)$.
The solution produces $(H_0,\Omega_{M,0})$ for $J$CDM (Fig. \ref{fig_BAO}). Table 1 lists the result alongside previous analysis in late-time cosmology by fits to tabulated $H(z)$ data and {\em Planck} $\Lambda$CDM analysis of the CMB. 
A previous analysis \citep{van21} produced (\ref{EQN_L3c}) as the solution to (\ref{EQN_ts}) and the age of the Universe according to {\em Planck}. The present BAO analysis leaves the latter unconstrained, replaced it by aforementioned {\em Planck} ratio of radiation-to-matter density for the case $g=1$.

\begin{figure*}
    \centerline{
    \includegraphics[scale=0.5]{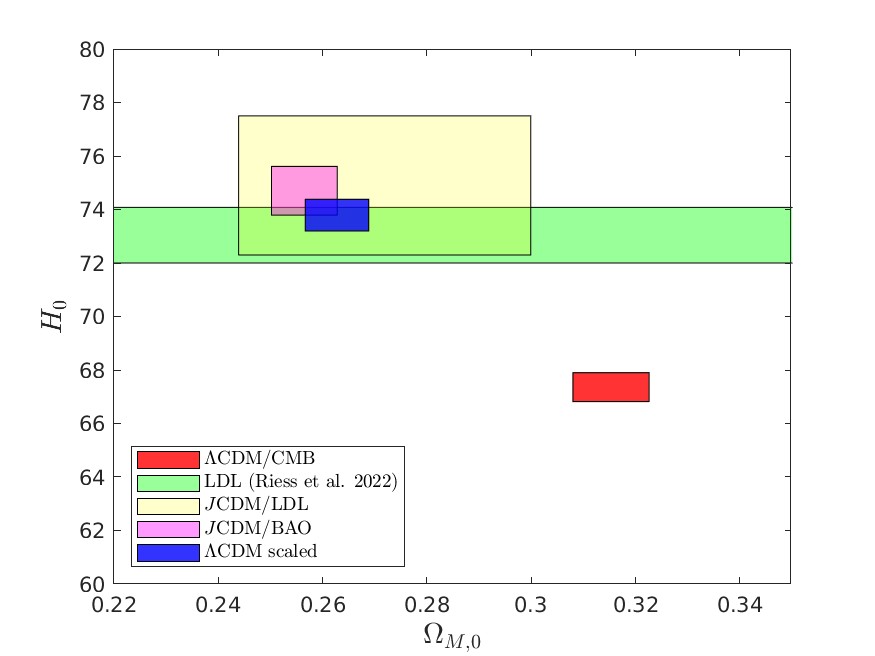}}
    \vskip-0.0in
    \caption{
    $J$CDM versus $\Lambda$CDM by {\em Planck} (red) and the LDL (green) in the $\left(H_0,\Omega_{M,0}\right)$ plane
    by independent fits to the LDL (green) and the BAO (\ref{EQN_ts}). Included are the $J$CDM values
    expected from scaling (\ref{EQN_scaling}) of the $\Lambda$CDM {\em Planck} values. (Reprtinted from \citep{van24d}.)}
    \label{fig_JCDM}
\end{figure*}

Table 1 confirms the anticipated scaling (\ref{EQN_scaling}) of the {\em Planck} $\Lambda$CDM values With no free parameters. $J$CDM for the first
time accounts for the $H_0$-tension in the late-time Universe based on early cosmology, constrained by the BAO (\ref{EQN_ts}). 
This scaling is accompanied with a reduced matter density, further alleviating tension in $S_8=\sigma_8\sqrt{\Omega_{M,0}/0.3}$, provided $\sigma_8$ is essentially unchanged between $\Lambda$CDM and $J$CDM.
Fig. \ref{fig_BAO} also shows $J$CDM preserving the astronomical age of the Universe $T_U=H_0^{-1}t_U$, where
  \begin{eqnarray}
     {t_U} = \int_{0}^\infty \frac{dz}{(1+z)h(z)};
     \label{EQN_tU}
 \end{eqnarray}
 in a previous analysis \citep{van21}, (\ref{EQN_tU}) was used as a constraint.

Fig. \ref{fig_JCDM} summarizes Table 1, highlighting 
the cosmological parameters in $J$CDM based on the
Local Distance Ladder, the BAO, and the
scaling relation (\ref{EQN_scaling}).

\section{Entropic constraint on variations of fundamental constants}

The JWST `Impossible galaxies' extend the existing tensions between the Local Distance Ladder and {\em Planck} to structure formation at cosmic dawn.
Before discussing this, the following disfavors relatively exotic solutions based on cosmic variations of fundamental constants.

Fundamental constants are the building blocks of our standard model of particle physics and gravitation. Specifically, this includes the Planck mass $m_p=\sqrt{\hbar c/G}$.
Dirac \cite{dir37} famously speculated on coincidences between the present age of the Universe relative to the atomic time-scale of the electron and the hierarchy problem, posed by the ratio of electric to gravitational interactions between the proton and the electron. These early considerations have been explored to account for some of the tensions in cosmological parameters by potential variations on a Hubble time scale, including the fine-structure constant \citep{bek09,uza03,uza11}.  
Notably, recently JWST observations reveal ultra high-$z$ galaxies \citep{coe13,oes16,eis23,aus23}, upending early galaxy formation predicted by $\Lambda$CDM \citep{agh20}. Perhaps this might be explained by cosmic variances of fundamental constants over a Hubble time scale \citep{gup22,gup23}?

By universality, a variation of fundamental constants carries far-reaching implications throughout, warranting careful consideration \citep{tel48}. 
Variation of Newton's constant $G$ is subject to multiple bounds from astronomy \citep[e.g.][]{gup22}, laboratory experiments \citep[e.g.][]{car12} and the propagation of gravitational waves \citep{jia23}. 

Black holes - vacuum solutions to (\ref{EQN_R}) - are the most compact objects in the Universe by the universal extremal density (\ref{EQN_rhoc}) and (\ref{EQN_TG4}) at gravitational radius $R_g=GM/c^2$ for a mass $M$.
They carry a Bekenstein-Hawking entropy $S=k_BN$, where $N=(1/4)N_p$ is 
one-fourth the horizon area $A_H=4\pi R_S^2$ in units of Planck area for a Schwarzschild black hole of radius $R_S=2R_g$.

Astrophysical black holes \citep{kor13,gen03,gen10,ghe03,ghe08,abb16,par22} 
carry an arrow of time by $S$, in addition to a primordial direction of time since the Big Bang \citep{pen79}. 
Indeed, $S$ is non-decreasing \citep{bek73,bek81,haw74,haw75} by the Hawking black hole-area theorem \citep{haw71} with recent observational support from binary black hole mergers \citep{isi21}. 
This poses a constraint on variations in the fundamental constants that define $S$ in light of the remarkably low photon emission rates \citep{bek95,spa16} 
$\nu\simeq40\left({M}/{M_\odot}\right)^{-1}{\rm s}^{-1}$ in Hawking radiation \citep{haw74,haw75}. 
Small primordial black holes may be an exception, that evaporate on time scales less than a Hubble time \citep[e.g.][]{vil21}.
 
 HST and JWST observe supermassive black holes at the center of galaxies \citep{kor13} throughout the observable Universe, now also soon after the Big Bang \citep{coe13,oes16}. 
 They carry a cosmic arrow of time by the second law of thermodynamics, 
\begin{eqnarray}
\delta S\ge0.
\label{EQN_S2}
\end{eqnarray}
A constraint on cosmic variations of aforementioned $m_p$ and $k_B$ follows in light of 
\begin{eqnarray} 
S \sim k_B \left(\frac{M}{m_p}\right)^{2} \sim \left(\frac{ k_BG}{\hbar c}\right) M^2.
\label{EQN_S}
\end{eqnarray}
Quite generally, (\ref{EQN_S2}-\ref{EQN_S}) constrain cosmic variations of $m_p$ \citep{lin22,lom19,sob21,ses13} and $k_B$ given by 
\begin{eqnarray}
\frac{\dot{m}_p}{m_p}\le \frac{\dot{k}_B}{2k_B}, 
\label{EQN_V1}
\end{eqnarray}
where the dot denotes differentiation with respect to time. 
The same can be expressed as a function of cosmological redshift $z$, 
taking into account $dz/dt = -(1+z)H$ given $H=\dot{a}/a > 0$.
By (\ref{EQN_S2}-\ref{EQN_S}), $S\sim Gk_B/\hbar c$, whereby variations of fundamental constants are subject to
\begin{eqnarray}
\frac{{G}^\prime}{G} \le \frac{{\hbar}^\prime}{\hbar} + \frac{{c}^\prime}{c} - \frac{{k}_B^\prime}{k_B},
\label{EQN_V2}
\end{eqnarray}
where the prime refers to differentiation with respect to $z$.

Based on (\ref{EQN_S2}), (\ref{EQN_V1}-\ref{EQN_V2}) applies to cosmology on a Hubble time, even when considering a generalized second law of thermodynamics \citep{bek73} in interaction with black hole environments. 
For instance, $M$ and $S$ diminish by energy and entropy by evaporation.
The black hole luminosity satisfies 
$L_H\sim L_0/n$, where $A_H=3n\log3\,l_p^2$ \citep{hod98}.
Evaporation extends over a number of Hubble times 
$N_{ev}=H_0t_e\simeq 7\left(M_\odot/m_p\right)\left(M/M_\odot\right)^{3}$
leaving $\Delta M/M\simeq 1/N_{ev} \simeq 10^{-38}$. 
$\Delta S/S\simeq 2\Delta M/M$ hereby does not change (\ref{EQN_S2}).

In a recent proposal \citep{gup23},
the JWST ultra high-$z$ galaxies \citep{eis23,aus23} might derive from variations of the fundamental constants, decaying by a factor of two over a Hubble time subject to $G\sim c^3 \sim \hbar^3 \sim k_B^{3/2}$ \citep{gup22}. 
The accompanying Planck mass $m_p \propto {1}/{\sqrt{c}}$ then increases with cosmic time. Combined with the proposed scaling $k_B\propto {c}^2$, black hole entropy scales accordingly as $S\sim c^{3}$. In violation of the second law of thermodynamics, $S$ is then {\em decreasing} by a decay in $c$ over a Hubble time.

Independent of the fine-structure constant of the electromagnetic interactions considered previously \citep{dir37}, black holes pose a novel constraint on cosmic variations of $m_p$ and $k_B$.
By (\ref{EQN_V2}), this rules out stretching time by covarying cosmic variations of fundamental constants. 

Alternatively, one might appeal to slow early expansion \citep{mel23}, allowing extended time for galaxy formation and more so than what is expected in conventional matter-dominated evolution $a(t)\propto t^{2/3}$.
However, this implies a vanishing deceleration parameter $q_0= -\ddot{a}a/\dot{a}^2\equiv 0$, which is ruled out by observations \citep[e.g.][]{abc21}. 

Instead, the JWST `Impossible galaxies' at ultra high-$z$ more likely point to rapid galaxy formation on a conventional matter-dominated cosmological background. 
To this end, we next turn to the origin of curvature in general relativity in information, based on quantum field theory.

\section{Curvature from consistent IR coupling to position information}

The simplest case of position information is binary: Schr\"odinger's cat is inside or outside a box.
This information is encoded on the box, derived from the amplitude of spacelike transitions across according to its propagator.
This encoding is not without limits. Expressed by the Bekenstein bound \citep{bek81}, the cat cannot be arbitrarily
overweight and still fit in the box of a given size. We formalize this here as follows \citep{van24e}.

\subsection{Position information and potential energy}

The propagator of a particle of mass $m$ at the Zitter angular frequency $\omega \hbar =mc^2$ in a cube of dimension $L$ defines a {\em position information} $I=12 \varphi$ by the Compton phase $\varphi=\omega s/c$ over $s=L/2$
\citep{van15a}.
(In fact, $I$ is constant regardless of the position within a cube.) 
For a concentric sphere of radius $s$, $I$ is encoded more efficiently, reduced by orthogonal projection of the area $24s^2$ of the cube to the area $4\pi s^2$ of the enclosed concentric sphere of radius $s$ by the area ratio $\pi/6$. 
We thus arrive at the dimensionless position information on the sphere \citep{van15a} (Fig. 4)
\begin{eqnarray}
    I = 2\pi \varphi .
    \label{EQN_I}
\end{eqnarray}

The expression (\ref{EQN_I}) is universal, applying also to the energy $\epsilon_\nu=\omega\hbar$ of radiation at the Unruh temperature $k_BT=\kappa \hbar/2\pi c = \hbar c/2\pi\xi$ of a Rindler horizon $h$ at distance $\xi=c^2/\kappa$, seen by an observer at acceleration $\kappa$. 
The Boltzmann factor $e^{-\hbar\omega/k_BT}$ of radiation at frequency $\omega$ has a corresponding phase jump $\varphi = \omega\xi/c$, whereby
${\hbar\omega}/{k_BT}=2\pi \varphi$ once more. It highlights position information (\ref{EQN_I}) to be an energy ratio which, for a massive particle, derives from its the Zitter frequency. As the exponent of a Boltzmann factor, 
\begin{eqnarray}
e^{-\hbar\omega/k_BT}=e^{-2\pi\varphi},
\label{EQN_exp}
\end{eqnarray}
$I$ hereby has an equivalent interpretation in entropy $S$ in the Boltzmann factor of (thermal) radiation, 
\begin{eqnarray}
 S= k_BI. 
 \label{EQN_SI}
\end{eqnarray}

\begin{figure*}
    \centering
    \includegraphics[scale=0.44]{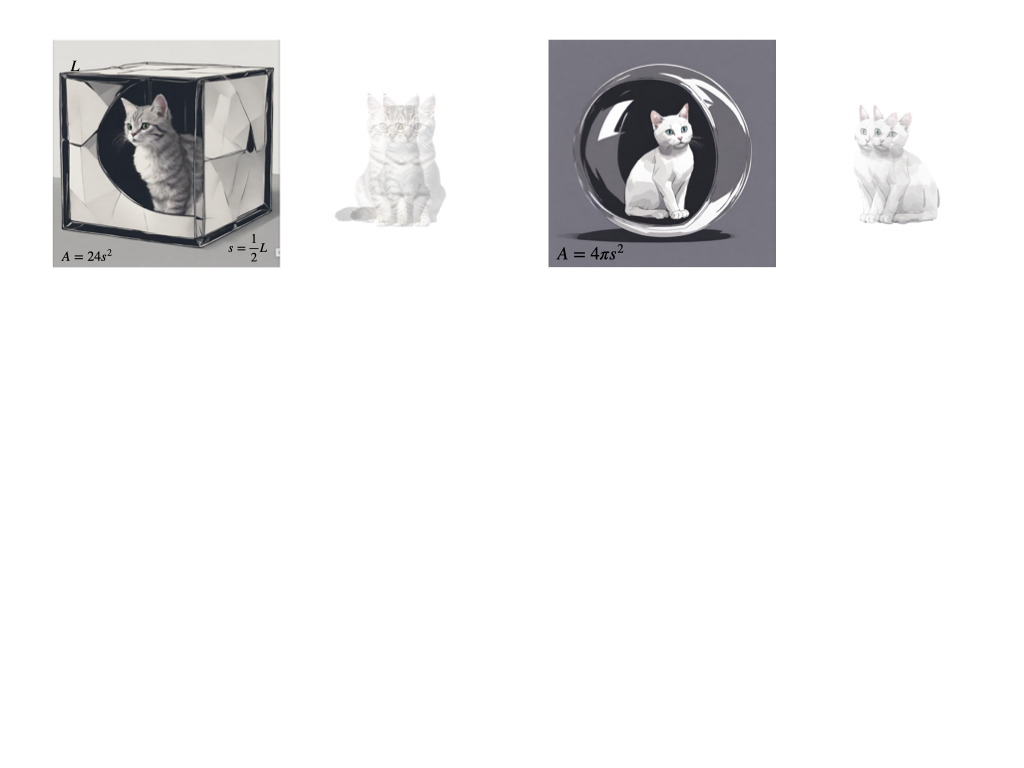}
    \vskip-2.9in
    \caption{
    By Compton phase $\varphi$, a cat in a box (left) has a finite probability to be outside according to the propagator of a massive particle. This information is encoded on the box according to the logarithm $I_0=12\varphi$. 
    Encoding on a sphere is more efficient according to the area ratio $\pi/6$ \citep{van15a}, leaving $I=2\pi\varphi$.
    As a UV-divergent quantity, $I\sim 1/\hbar$, is a primitive that calls for a IR consistent coupling to spacetime in classical general relativity.
    (Reprinted from \cite{van24f}.)}
    \label{FIGphi}
\end{figure*}


Relative to a horizon $h$, consider a particle at a distance $\xi$ has $I=2\pi\varphi$ and entropy (\ref{EQN_SI}).
The temperature of the Unruh vacuum is a thermodynamic temperature 
\begin{eqnarray}
    k_BT_U=\left(\frac{\partial S}{\partial E}\right)^{-1} 
    = \frac{\hbar c}{2\pi \xi}, 
    \label{EQN_Uh1}
\end{eqnarray}
where $E=mc^2$. 
In the gravitational field of $h$, the relative 
potential energy $U_h$ derives as an entropic force \citep[cf.][]{ver11}
\begin{eqnarray}
    F = T_\kappa \frac{\partial S}{\partial s}
    = \left(\frac{\hbar c}{2\pi \xi}\right)
    \left(2\pi \frac{mc}{\hbar}\right) 
    = \frac{mc^2}{\xi}
    \label{EQN_Uh3}
\end{eqnarray}
pointing to $h$.
Specializing to $h$ in Rindler space, 
$\kappa=c^2/\xi$ reduces (\ref{EQN_Uh3})
to Newton's law $F=m\kappa$ in 
\begin{eqnarray}
    U_h = \int_0^\xi Fds = m\kappa \xi = mc^2.
    \label{EQN_Uh2}
\end{eqnarray}
In the above, (\ref{EQN_Uh2}) identifies $U_h$ with
the work performed in extracting an amount of heat
$Q=mc^2$ from $h$.
To go beyond, including $G$, we consider the following.

\subsection{Bekenstein-Hawking entropy}

In spherical symmetry, $I$ has a lower bound given by the Bekenstein-Hawking entropy. To derive this from Rindler spacetime, we appeal to the equivalence principle. By Planck length $l_p$, this introduces $G$ in addition to $c$ and $\hbar$.

Encoding (\ref{EQN_I}) on a sphere of radius $r$ for a gravitating mass $M$ with gravitational radius $R_g=GM/c^2$ assumes an Einstein area  
\begin{eqnarray}
A_E=\lambda_3 I l_p^2,
\label{EQN_AE}
\end{eqnarray}
where the constant $\lambda_3$ derives from the equivalence principle as follows.
Consider $M$ as the {\em Man-in-the middle} in the strip of width $\xi = c^2/\kappa$ in a Rindler frame ${\cal R}$ at acceleration $\kappa=GM/r^2$ seen by a test particle $m$. $M$ is between the Rindler horizon $h$ and $m$.
A distant observer holding on to $m$ infers $\xi=r^2/R_g$, where $\xi \ge 2r$ by causality. 
The limit $\xi=2r$ holds when $m$ reaches infinite redshift of $h$, identifying antipodal points
by virtue of spherical symmetry about $M$. 
$M$ hereby possesses a surface $H$ of infinite redshift of Schwarzschild radius $R_S$. 
To finalize, $dI=2\pi d\varphi$, $d\varphi = \xi dR_g/l_p^2$, of a mass element $dM$ relative to $h$, whereby
\citep{van24c}
\begin{eqnarray}
    S=k_B l_p^{-2}\int_0^M \xi dR_g 
    = 4\pi k_B R_g^2/l_p^2 = \frac{1}{4}k_BN_p, 
    \label{EQN_14}
\end{eqnarray}
where the entropy $S$ is interpreted as $k_BI$ hidden behind the event horizon $H$ of area $N_p=A_H/l_p^2$ in
Planck units in the notation of (\ref{EQN_IR}).
Crucially, (\ref{EQN_14}) identifies the Bekenstein-Hawking entropy formula {\em sans reference to thermodynamics}. 

In (\ref{EQN_AE}), $\lambda_3=4$ by (\ref{EQN_14}) will be assumed to be universal. Extended to $r\ge R_S$, $R_S=2R_g$, independent of redshift, this assumption recovers (\ref{EQN_I}) from the above-mentioned derivation in Rindler spacetime and the equivalence principle.

\subsection{Curvature from Huygens' principle}

The encoding (\ref{EQN_I}) in $A_E$ (\ref{EQN_AE}) affects the propagation - and hence the distribution - in $\varphi$ about $m$. $\varphi$ is a scalar field, which provides a covariant radial distance in spherically symmetric wavefronts. 
By $I\sim1/\hbar$, position information (\ref{EQN_I}) is a second primitive. 
A consistent IR coupling to general relativity derives on concentric surfaces surface $A_\varphi$ of 
constant phase $\varphi$ of area $A=A_pl_p^2 = 4\pi r^2$ by the surface density 
\begin{eqnarray}
p= \frac{A_E}{4\pi r^2} = 8\pi \varphi \alpha_p
\label{EQN_p1}
\end{eqnarray}
of entanglement with $M$ according to (\ref{EQN_I}), where $\alpha_pA_p=1$ conform (\ref{EQN_IR}).
The result $p=2u$ in (\ref{EQN_p1}) shows {\em the Newtonian gravitational potential $u=R_g/r$
is the primitive $I$ with consistent IR coupling to general relativity.}

A remainder $1-p$ partakes in wave propagation according to Huygens' principle, leaving a velocity
\begin{eqnarray}
 \beta = 1- p,
 \label{EQN_p}
\end{eqnarray}
where $\beta = v/c$ denotes the three-velocity relative to the velocity of light. 
In practical terms, $\beta$ describes the remaining window (by surface fraction) through which  
the proverbial cat of mass $m$ can escape by tunneling to the outside.
The limit $\beta=0$ in (\ref{EQN_p}) for $p=1$ in (\ref{EQN_p1}) attains for $A_E=4\pi r^2$,
when $A_\varphi$ has the Schwarzschild radius of $m$, while $\beta\simeq 1$ at large distances relative to the same.
This radius dependent velocity $\beta$ in (\ref{EQN_p}), probed by $I\alpha_p$ as a massless scalar field indicates 
spacetime curvature based on wave motion - a distance measured by total phase $\varphi$ of a spherical 
wave about a point mass \citep{van12}.

In Minkowski spacetime, we have $j=A_0/A_\varphi=1/4$ for the ratio of poloidal surface area $A_0=\pi r^2$
to spherical surface area $A_\varphi=4\pi r^2$. 
More generally, $\beta = dr/cdt$ satisfies $\beta =\alpha^2$ in terms of the
redshift factor $\alpha$ in a spherically symmetric line-element $ds^2=\alpha^2 c^2dt^2 + dr^2/\alpha^2$ 
of radial wave motion of a massless field in curved spacetime. According (\ref{EQN_p}) and
$A_0=\int^r 2\pi r ds$, $ds=dr/\alpha$, $A_0 \simeq \pi r^2 + I l_p^2$ in this gauge 
$A_\varphi=4\pi r^2$, whereby
\begin{eqnarray}
    j \simeq \frac{1}{4} + \frac{I}{A_p}.
    \label{EQN_j}
\end{eqnarray}
Changing gauge to $A^\prime = \pi r^2$, (\ref{EQN_j}) explicitly identifies 
$A_\varphi=4\pi r^2 -A_E$ 
with {\em regressed propagation} (\ref{EQN_p}).

By (\ref{EQN_p1}-\ref{EQN_j}), curvature in general relativity 
represents the consistent IR coupling to position information of $M$ in (\ref{EQN_p1}). 
The Schwarzschild limit of (\ref{EQN_p1}-\ref{EQN_p}) corresponds to
$\varphi=kr$, $k=mc/\hbar$, at $r=2R_g$ of $m$, where $R_g=GM/c^2$.
We next consider the implications on a cosmological background,

\section{Galaxy dynamics tracing background cosmology}

Galaxy dynamics offers a setting to study the limit of weak gravitation at the lowest accelerations in the Universe. To this end, consider a particle in orbital motion with centripetal acceleration $\alpha$ at a radius $r$ in a spiral galaxy of mass $M$. We determine $\alpha=V_c^2/r$ according to circular velocity $V_c$ by spectroscopy. 
For a particle of Newtonian inertia $m$ in the gravitational field $a_N=GM/r^2$ of $M$, the centripetal force represents Newton's second law $F_N=ma_N$. Allowing for the possibility of a non-Newtonian inertia $m^\prime$, momentum conservation requires 
\begin{eqnarray}
    \alpha m^\prime = ma_N.
    \label{EQN_alpha}
\end{eqnarray} 
Accordingly, gravitational binding energy $U_N=-\int_r^\infty F_N ds = -GM/r$ is preserved. In terms of the expected circular velocity $V_N$, $a_N=V_N^2/r$, $U_N$ is accompanied by the Newtonian kinetic energy $E_k=\frac{1}{2}mV_N^2$. 
By $V_c$, the kinetic energy $E_k^\prime$ satisfies
\begin{eqnarray}
E_k^\prime = \frac{1}{2} m^\prime V_c^{ 2} = \frac{1}{2} m V_c^{ 2} \left( \frac{m^\prime}{m}\right) =  \frac{1}{2} m V_c^{2} \left(\frac{a_N}{\alpha}\right) = \frac{1}{2} m V_N^2 = E_k.
\end{eqnarray}
As a result, the Hamiltonian 
\begin{eqnarray}
    H = E_k + U_N
    \label{EQN_Ha}
\end{eqnarray}
is {\em invariant under a change to non-Newtonian inertia}.

Indeed, it has been speculated that galaxy dynamics might reveal a finite sensitivity to the cosmological background \citep{san90}. 
To consider this, (\ref{EQN_Uh2}-\ref{EQN_Uh3}) provides a natural starting point for inertia in potential energy relative to a horizon surface that, by (\ref{EQN_SI}), identifies a thermodynamic origin in the background spacetime.
This points to reduced inertia by an IR cut-off in the integral (\ref{EQN_Uh3}) at crossing of Rindler and Hubble horizons \citep{van17}. Since acceleration is of dimension 1/cm and the trace of the Schouten tensor $\left[J\right]={\rm cm}^{-2}$ (in geometrical units $G=c=1$), such sensitivity may arise with $\sqrt{J}$. At finite sensitivity, Newton's second law is no longer scale free, but assumes a scale inhereted from background spacetime.

To model this, consider a particle in orbit of a galaxy of mass $M$ at aforementioned centripetal accelertion $\alpha$.
In the comoving Rindler spacetime, it experiences a gravitational field $g=-\alpha$ by the equivalence principle.
(Hence, $g+a_N=0$ in Newton's interpretation of orbital motion in free fall et semper.)
In this frame, the particle has a potential energy
$U_h=E$ relative to the Rindler horizon $h$, equal to its invariant rest-mass energy $E=mc^2$ by the potential energy $U_h$ to $h$ (\ref{EQN_Uh2}-\ref{EQN_Uh3}).
The Newtonian limit of constant inertia holds when $\xi < R_H$ on a cosmological background with Hubble radius $R_H$.

More generally, the potential energy $U$ to the {\em nearest} event horizon satisfies
\begin{eqnarray}
 U\le E,
\label{EQN_U1}
\end{eqnarray}
Across the transition radius (\ref{EQN_rt}), $a_N < a_{dS}$ by $\xi > R_H$, however. By causality, 
${\cal H}$ cuts aforementioned strip to $R_H<\xi$.
According to (\ref{EQN_Uh2}-\ref{EQN_Uh3}), this leaves
\begin{eqnarray}
    U\sim \frac{R_H}{\xi}U_h.
    \label{EQN_U2}
\end{eqnarray}
Inertia in $U$ is hereby reduced by a factor $R_H/\xi$,
as Newton's second law assumes a scale $a_{dS}$, equivalently, $R_H$, in the application of the equivalence principle to background cosmology. 

\begin{figure*}
    \centering
    \includegraphics[scale=1.3]{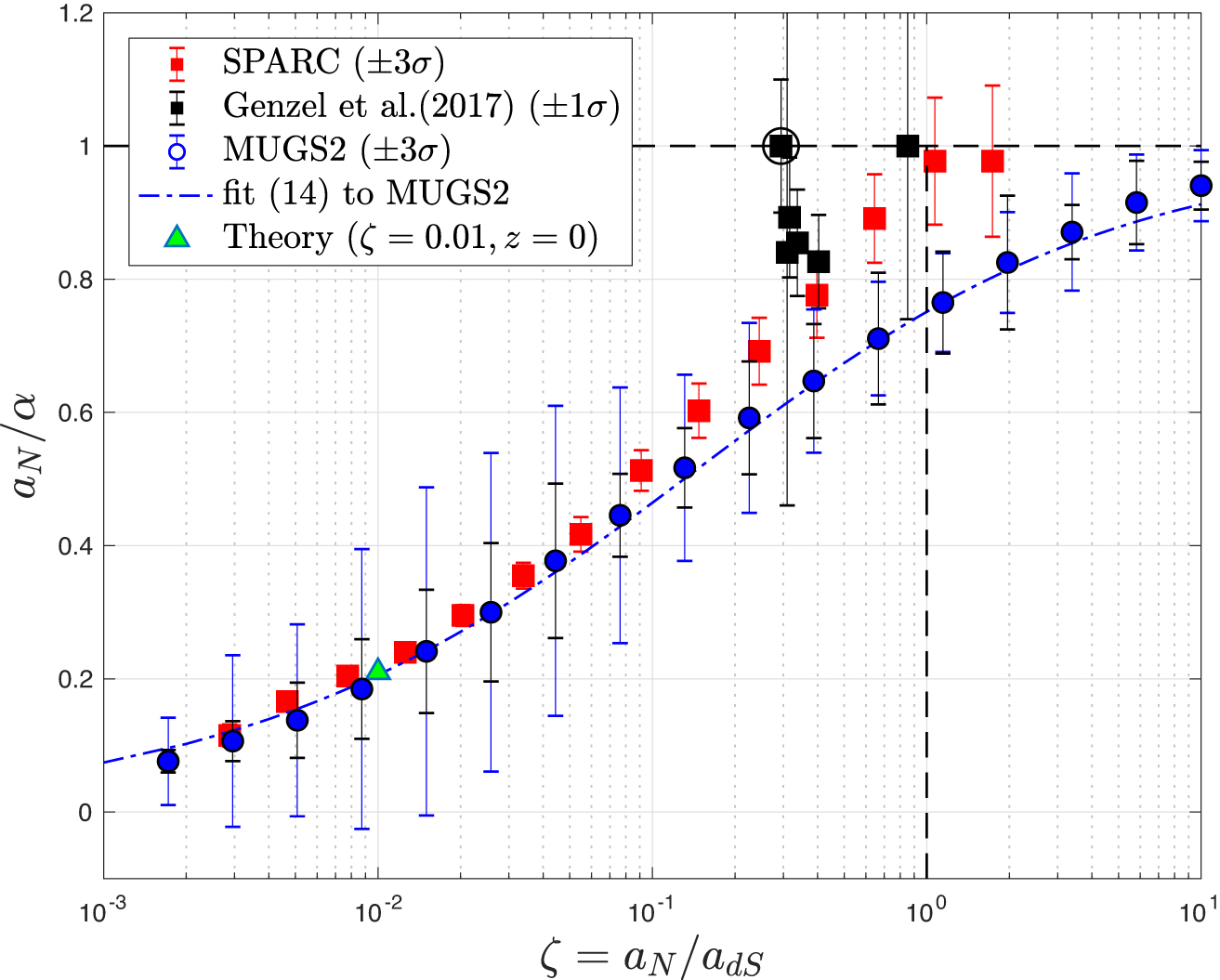}
    \vskip-0.0in
    \caption{
    A normalized plot of rotation curves of SPARC versus $\Lambda$CDM galaxy models. Shown are observed radial accelerations $\alpha = V_c^2/r$ versus expected Newtonian radial accelerations $a_N$, inferred from circular velocities $V_c$ at radius $r$ and baryonic mass. Plotted is $a_N/\alpha$ versus $\zeta$: $a_N$ normalized to the de Sitter acceleration $a_{dS}$ of the background cosmology. 
    SPARC (red) shows a sharp $C^0$-transition across the de Sitter scale of acceleration $a_{dS}$ between Newtonian and anomalous dynamics in galaxy rotation curves. This normalized plot shows a 6$\sigma$ gap relative to $\Lambda$CDM galaxy models in $\Lambda$CDM (blue). It signifies the absence of conventional CDM. Instead, this points to low-energy dark matter with de Broglie wavelengths on the scale of 
    $r_t$ (\ref{EQN_rt}) or larger, rendering dark matter inherently transparent to standard model interactions.
    The green triangle represents (\ref{EQN_a00})
    predicted by (\ref{EQN_a0}).
    [Reprinted from \cite{van18}.) 
    }
    \label{figRC}
\end{figure*}

For a galaxy of mass $M=M_{11}10^{11}M_\odot$, the radial acceleration $\kappa=V_c^2/r$ at circular velocity $V_c$ carries a crossing $\xi\simeq R_H$ across the de Sitter scale of acceleration (\ref{EQN_adS}). 
By (\ref{EQN_U1}-\ref{EQN_U2}), this predicts an essentially $C^0$-transition in galaxy dynamics 
across $a_N  = a_{dS}$. For a galaxy of gravitational radius $R_G=GM/c^2$, this is across a corresponding transition radius
\citep{van17}
\begin{eqnarray}
    r_t = \sqrt{R_HR_G}= 4.7\,M_{11}^{1/2}{\rm kpc}
    \label{EQN_rt}
\end{eqnarray}
at the current epoch $z=0$. (\ref{EQN_rt}) bears out clearly in the normalized plot Fig. \ref{figRC}, showing the significance of $a_{dS}$ in the location of a sharp transition between Newtonian and anomalous galaxy dynamics.

To describe the asymptotic region $a_N\ll a_{dS}$ corresponding to $r\ll r_t$, we note the following.

\subsection{Inertia on a cosmological background}

On a cosmological background, propagation of a wavefront $A_\varphi$ considered in \S6.1 will be subject to dispersion according to the dark energy density $J$.
Such is consequential for inertia as a thermodynamic quantity based on position information at finite temperature \citep{van17c}. In what follows, we derive this based on a minimal extension of (\ref{EQN_p1}) by the mass-less scalar field $p\rightarrow pe^{i\varphi}$. As a fraction of
area, it satisfies the conformally invariant Klein-Gordon equation. In four-dimensional spacetime, it picks up dispersion  
\begin{eqnarray}
\omega = c\sqrt{k^2 + J}
\label{EQN_disp1}
\end{eqnarray}
between angular frequency $\omega$ and wave number $k$, where the $J$ implies a minimum mass $\hbar\sqrt{J}$.
In applying (\ref{EQN_Uh2}-\ref{EQN_Uh3}), we shall consider
\begin{eqnarray}
    \partial_a \varphi = N\left(\omega/c,k\right)
    \label{EQN_disp2}
\end{eqnarray}
where $N$ refers to the number of wave modes involved in the encoding of the energy and position of a particle.
At a distance $r$ of a gravitating mass $M$, $R_g=GM/c^2$, is subject to a centripetal Newtonian gravitational acceleration $a_N=GM/r^2$. In the comoving frame of reference, 
$\kappa=a_N$ formally defines an associated Rindler horizon $h$ at a distance $\xi$. In the Newtonian limit, 
the 2-sphere $A_\varphi$ about $M$
covers $N=I$ Planck-sized degrees of freedom satisfying $2Nk_BT_\kappa=mc^2$, where $I=2\pi\varphi$ 
is evaluated invariantly by $\varphi = \omega r/c$ or $\varphi=kr$. By (\ref{EQN_disp2}), however, $\omega\ne kc$, breaking this Newtonian correspondence.
At the same time, the total vacuum temperature 
\begin{eqnarray}
T=\sqrt{T_H^2+T_\kappa^2}
\label{EQN_T}
\end{eqnarray}
satisfies \citep{nar96,des97,jac98,kli11}.
Here, $T_H$ satisfies (\ref{EQN_TH}) whereby
(\ref{EQN_disp2} and (\ref{EQN_T}) introduce {\em two}
similar but not identical dispersion relations on the
background \citep{van17}. (They agree only in the unphysical limit $q=-3$.)

We next proceed with balance of energy and momentum.  
\begin{itemize}
\item
Subject to (\ref{EQN_T}), on the 2-sphere $A_\varphi$ about $M$, we have \citep[cf.][]{ver11}
\begin{eqnarray}
   mc^2\simeq 2Nk_BT.
   \label{EQN_N2}
\end{eqnarray}
In the weak field limit relevant to the baryonic Tully-Fisher relation \citep[bTFR,][]{mcc12}, the number of Planck sized area elements partaking in (\ref{EQN_N2}) satisfies $N< I$ when $T_H> T_\kappa$ and $T$ effectively reduces to $T_H$ (\ref{EQN_TH}) at $\kappa \ll a_{dS}$. 
\item 
Given an acceleration $\kappa$, inertia is identified with potential energy $U$ (\ref{EQN_U2}) in Rindler space. According to (\ref{EQN_Uh2}-\ref{EQN_Uh3}), $m^\prime$ satisfies
\begin{eqnarray}
    m^\prime c^2 = \left(\frac{\hbar \kappa}{c}\right) \varphi.
    \label{EQN_IB}
\end{eqnarray}
Evaluated over the light-crossing time of the distance to the nearest horizon, 
\begin{eqnarray}
\varphi = N\min\left\{\xi,R_H\right\}\omega/c \simeq N \sqrt{J}R_H/c,
\label{EQN_IC}
\end{eqnarray}
where $\xi=c^2/\kappa$.
The right hand-side refers to the IR cut-off by $R_H\ll \xi$ in (\ref{EQN_U2}) in the potential energy (\ref{EQN_Uh2}-\ref{EQN_Uh3}), summing over the $N$ modes in (\ref{EQN_N2}) with dispersion (\ref{EQN_disp1}).
\end{itemize}
In weak gravitation $\kappa\ll a_{dS}$, (\ref{EQN_IB}-\ref{EQN_IC}) gives 
\begin{eqnarray}
    m^\prime c^2  
    = \left(\frac{\hbar c}{\xi}\right)\varphi = N\hbar\sqrt{J}\left(\frac{R_H}{\xi}\right).
    \label{EQN_N2c}
\end{eqnarray}
Combined with momentum conservation (\ref{EQN_alpha}) and (\ref{EQN_N2}), (\ref{EQN_N2c})
implies the observed acceleration $\alpha$ satisfies
\begin{eqnarray}
    \frac{a_N}{\alpha} = \frac{m^\prime}{m} = \frac{\sqrt{J}\hbar}{2k_BT_H}\left(\frac{R_H}{\xi}\right) = \frac{2\pi}{\sqrt{1-q}} \frac{\alpha}{a_{dS}}.
    \label{EQN_N2d}
\end{eqnarray}
In this ratio, $N$ in (\ref{EQN_N2}) and (\ref{EQN_N2c}) cancels, and the right-hand side follows from (\ref{EQN_TH}).
The Milgrom parameter $a_0$ in $\alpha = \sqrt{a_0a_N}$ \citep{mil84} hereby evolves on a cosmological background according to
\citep{van24e} 
\begin{eqnarray}
    a_0 = \frac{1}{2\pi}\sqrt{J}c^2
    \label{EQN_a0}
\end{eqnarray}
anticipated by above-mentioned dimensional analysis. 

\subsection{Weak gravitation in spiral galaxies}

The transition radius (\ref{EQN_rt}) is hereby distinctly smaller than the critical radius $r_c=\sqrt{R_G/a_0}c\simeq \sqrt{2\pi} r_t$
discussed previously \citep{san90}. Notably, therefore, the problem of anomalous dynamics in rotation curves is parameterized by the 
{\em two} acceleration parameters $\left(a_{dS},a_0\right)$.
The relation (\ref{EQN_a0}) makes explicit earlier suggestions for a connection between (reduced) inertia and the cosmological vacuum \citep{mil99,san90}, allowing a study of its observational consequences over a wide range of redshifts.

A suitable fit to the radial accelerations in the MUGS2 $\Lambda$CDM galaxy model is
\begin{eqnarray}
    \frac{a_N}{\alpha} \simeq \left(\frac{1}{2}+\sqrt{\frac{1}{4}+\frac{1}{x}+\frac{1}{\sqrt{x}}}\right)^{-1}
    \label{EQN_fit}
\end{eqnarray}
as a function of $x=4\pi \zeta$, $\zeta=a_N/a_{dS}$. The fit (\ref{EQN_fit}) satisfies $a_N/\alpha = 0.78$ at $\zeta=1$ - a gap of 22\%
below the observed $a_N/\alpha\simeq 1$ at $\zeta=1$. This gap was missed in the original analysis of SPARC versus MUGS2 $\Lambda$CDM galaxy models \citep{kel17}.

Asymptotically far out in a galactic disk, we have
\begin{eqnarray}
    \frac{a_N}{\alpha} = \sqrt{2\pi}\left(1-q\right)^{-1/4}\zeta^{1/2}\simeq 2.1\,\zeta^{1/2},
    \label{EQN_a00}
\end{eqnarray}
indicated by the green triangle in Fig. \ref{figRC}.

By (\ref{EQN_a0}) and the bTFR $M_b=AV_{max}^4$ correlating baryonic galaxy mass with maximal rotation velocities $V_{max}$ in spiral galaxies, we infer \citep{van24}
\begin{eqnarray}
q_0 = 1- \left( \frac{2\pi}{GA a_{dS}}\right)^2 \simeq -0.98^{0.5}_{0.5},
\label{EQN_q0}
\end{eqnarray}
where $A=\left(47\pm 6\right)M_\odot$km$^{4}$s$^{-4}$ \citep{mcc12}.
The $q_0$-estimate (\ref{EQN_q0}) is consistent and independent of the $q_0$-estimate based on the Local Distance Ladder \citep{cam20}. 

\subsection{Early galaxy formation at cosmic dawn}

The JWST `Impossible galaxies' discussed in \S5 upend $\Lambda$CDM by light at cosmic dawn about an order of magnitude earlier than expected. It indicates rapid formation of the first stars and galaxies, beyond gravitational collapse governed
by Newtonian theory on the early expansion history $a\propto t^{2/3}$, a background essentially fixed by tight 
observational constraints, notably by the astronomical age of the Universe and the BAO in the surface of last scattering (SLS).

Galaxies originate from primordial density fluctuations in the SLS by gravitational collapse. 
We can infer time-scales of gravitational collapse from the 2-body problem. According to (\ref{EQN_a0}), 
collapse times $t_c$ in weak gravitation below the de Sitter scale of acceleration are shorter than collapse times $\tau_c$ in Newton's theory. Galaxy formation speeds up by accelerated gravitational collapse times according to the ratio $\zeta = a_N/a_{dS}$, satisfying \citep{van24}
\begin{eqnarray}
    \frac{t_c}{\tau_c}=0.94\left(\frac{a_0}{a_{N,0}}\right)^{1/4}=1.62\zeta^{-1/4} \simeq 25 M_9^{-\frac{1}{12}}\left(1+z\right)^{1/8},
    \label{EQN_FF}
\end{eqnarray}
leaving the initial distribution of galaxy masses $M=M_910^9M_\odot$ essentially unchanged.
Crucially, this accounts for the JWST `Impossible galaxies'  by the same first principles giving rise to
the bTFR with no free parameters. 

\section{Conclusions and outlook} 

We present a $J$CDM cosmology beyond $\Lambda$CDM with no free parameters derived from a Big Bang quantum cosmology with primitives $\Lambda_0$ and $I$.
Consistent IR coupling by (\ref{EQN_IR}) to spacetime gives rise to a dark energy (\S2) and, respectively, curvature embedding the Newtonian gravitational potential (\S6).

This recognizes $\Lambda_0$ to be essential to the identification $\Lambda=J$ with the trace $J$ of the Schouten tensor (\ref{EQN_L1}). It formalizes de Sitter heat (\ref{EQN_Q}), quite distinct from a crisis in cosmology \cite{wei89}.

With no free parameters, $J$CDM predicts a scaling  of the Hubble constant (\ref{EQN_scaling}). Crucially, this accounts for the existing tension between the Local Distance Ladder and early cosmology according to the BAO in the {\em Planck} $\Lambda$CDM-analysis of the CMB (Table 1).

On sub-horizon scales, $J$CDM points to ultra-light dark matter - a cosmological distribution of essentially transparent matter of by wavelength(s) exceeding the transition radius (\ref{EQN_rt}) between Newtonian and anomalous galaxy dynamics 
across the background de Sitter scale of acceleration $a_{dS}$ (\ref{EQN_adS}). Preserving the $C^0$-transition across $a_{dS}$, predicted and observed in SPARC (Fig. \ref{figRC}) puts an upper bound on the mass of this putative dark matter beyond the standard model of particle physics. 

$J$ unifies dark energy and anomalous galaxy dynamics
in (\ref{EQN_L1}), (\ref{EQN_a0}) and (\ref{EQN_a00}). While general relativity is preserved \citep{wei22}, it clearly receives contributions from its UV-completion in quantum gravity.

\subsection{$J$CDM and its observational consequences}

$J$CDM introduces a cosmological vacuum with dynamical and non-local dark energy. On this background, bTFR and early galaxy formation are unified by a finite sensitivity to $\sqrt{J}$ in weak gravitation.

In $J$CDM, the propagation of gravitational and electromagnetic waves satisfy the same dispersion relation, explicitly in their respective formulations in SO(3,1) and U(1) in the Lorenz gauge \citep{van96}.
Consistency is observed within one part par $10^{-15}$ given the small time-delay of 1.7\,s between GW170817 and GRB170817A \citep{abb17}. In fact, the time delay of 0.78\,s between GW170817B and GRB170817A \citep{van23} tightens this observational constraint by a factor of about two.

\mbox{}\\
\centerline{ {\em (a) The vacuum of a Big Bang cosmology}}
\vskip0.1in

The relic of the Big Bang (\ref{EQN_Q}-\ref{EQN_rhoc}) satisfies the canonical scaling of energy density $\rho_c\sim R_H^{-2}$ of black hole spacetimes by 
$\Lambda$ derived from (\ref{EQN_IR}).
$\Lambda$ is hereby dynamical, precluding the de Sitter solution of constant Hubble parameter. This rules out $\Lambda$CDM - with fozen $\Lambda$ - in the distant future, inevitably in tensions with the Local Distance Ladder. 
A formal derivation of $\Lambda$ derived from gauging global phase in the path integral formulation further shows $\Lambda=J$. $\Lambda$ hereby gives rise to
$H(z)$ and a deceleration parameter $q(z)$ of the background cosmology distinct from $\Lambda$CDM.

\mbox{}\\
\centerline{ {\em (b) Tension-free Hubble expansion}}
\vskip0.1in

While n$\Lambda$CDM assumes $\Lambda$ to be frozen, 
$J$CDM describes a dynamical dark energy of a vacuum at finite temperature (\S3). It satisfies a T-duality in the Friedmann scale factor $a$ with observational consequences for tensions in cosmological parameters expressed in (\ref{EQN_scaling}) of \S4.

\begin{figure*}
    \centering
    \includegraphics[scale=0.45]{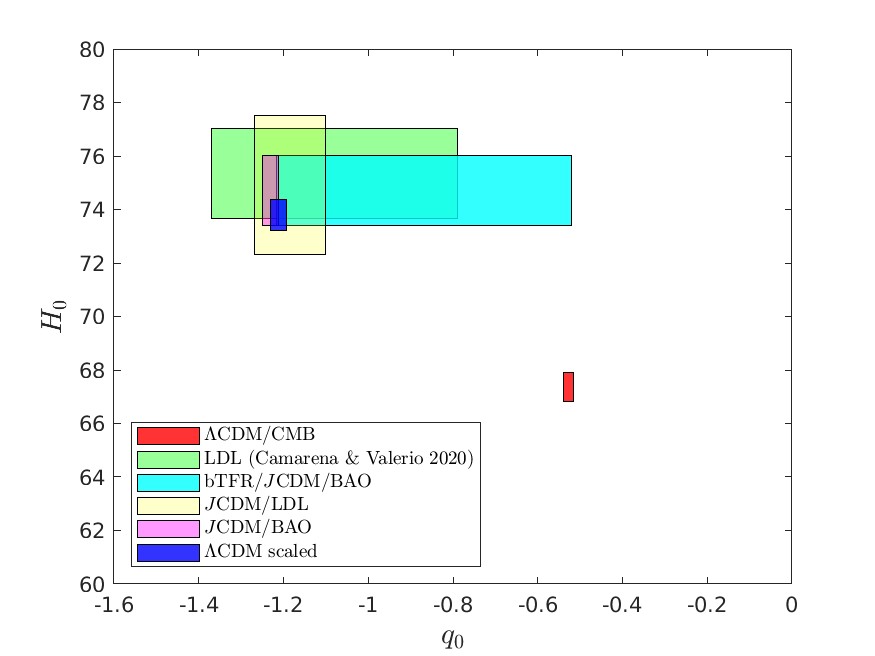}
    \vskip-0.0in
    \caption{
    Overview of $\left(H_0,q_0\right)$-tension in $\Lambda$CDM by {\em Planck} with respect to the LDL (green) and $J$CDM, where $J$CDM results are shown independently from the LDL (yellow) and the BAO (magenta). 
    Included are $J$CDM results from scaling (\ref{EQN_scaling}) of {\em Planck} values.
    (Reprinted from \citep{van24d}.)
    }
    \label{figH0q0}
\end{figure*}

While $J$CDM and $\Lambda$CDM share essentially the same Hubble expansion in the past, $J$CDM does not approach  de Sitter in the distant future allowing consistency with the Local Distance Ladder. 
Tensions with $\Lambda$CDM are identified with scaling of {\em Planck} values (\ref{EQN_scaling}), 
confirmed by solving the BAO angle as a constraint at recombination (Fig. \ref{fig_BAO}).
By (\ref{EQN_scaling}), $J$CDM further predicts 
\begin{eqnarray}
q_{0}=\Omega_{M,0}-2\Omega_{\Lambda,0}\simeq \frac{5}{3}q_0^\Lambda - \frac{1}{3} =-\frac{7}{6}\simeq -1.16, 
\label{EQN_q0J}
\end{eqnarray}
consistent with the Local Distance Ladder (Table 1, Fig. \ref{figH0q0}) distinct from $q_{0}^\Lambda =\frac{1}{2}\Omega_{M,0}-\Omega_{\Lambda,0}\simeq -0.5$ of $\Lambda$CDM.

In the outcome (\ref{EQN_q0J}) the contribution $J$ is absorbed non-locally in the Einstein tensor (\ref{EQN_H0}). The Hamiltonian energy constraint on $a(t)$ - the first Friedmann equation - is hereby {\em second order} in time, distinct from the first order Hamiltonian energy constraint in $\Lambda$CDM.
{\em $\Lambda$CDM is a singular limit upon ignoring a non-local contribution due to breaking time-translation invariance on a Hubble time.}
The main tensions between $J$CDM and $\Lambda$CDM in (\ref{EQN_scaling}) conceivably also alleviates $S_8$.

\mbox{}\\
\centerline{ {\em (c) Galaxy dynamics tracing background cosmology}}
\vskip0.1in

A finite sensitivity of galaxy dynamics to $\sqrt{J}$ in weak gravitation below $a_{dS}$ carries observational consequences across all redshifts. Notably, this applies to the bTFR and galaxy formation at cosmic dawn, recently highlighted by the JWST 'Impossible galaxies'.

bTFR reveals anomalous galaxy dynamics in weak gravitation below the de Sitter scale of acceleration $a_{dS}$, distinct from Kepler's laws in Newton's model of the solar system. Indeed, SPARC rotation curves (Fig. \ref{figRC}) identify an essentially $C^0$-transition across $a_{dS}$, evidenced by a $6\sigma$ gap from MUGS2 $\Lambda$CDM-galaxy models \citep{van18}. This gap unlikely originates in conventional CDM, such being inherently smooth due to diffusion by scattering canonical to $N$-body systems. The apparent $C^0$-transition allows for ultra-light DM with wavelengths $\lambda$ on the scale of 
galaxies \citep{van17,van18}, a cosmological distribution of which is required to preserve three-flatness, when the $C^0$-transition is accounted for by anomalous inertia in a departure from Newton's second law. 

Preserving the $C^0$-transition due to reduced inertia requires a de Broglie wavelength
$\lambda > r_t$. For CDM, this implies \cite{van24d}
\begin{eqnarray}
    m_Dc^2 < \max_\beta\left(\pi\frac{\hbar c}{r_t\beta}\right)
    \simeq 3\times 10^{-21}{\rm eV}
    \left(\frac{M}{10^{11}M_\odot}\right)^{-1/2}\left(\frac{H_0}{73\,{\rm km\,s}^{-1}{\rm Mpc}^{-1}}\right)^{1/2},
    \label{EQN_mD}
\end{eqnarray}
where $\beta=v/c$ for the reference value $r_t=5{\rm kpc}$ for a Milky Way type galaxy. Here, $\beta$ is bounded below by a crossing time $r_t/(\beta c)\lesssim T_U$ less than the age of the Universe $T_U$, i.e., $10^{-6}<\beta \ll 1$.  
The right-hand side of (\ref{EQN_mD}) is attained at $\beta\simeq10^{-6}$ with corresponding dark matter temperature
$k_BT_{DM}\sim k_BT_{dS}\simeq 5\times 10^{-33}$eV. At these temperatures, (\ref{EQN_mD}) naturally forms a condensate with critical temperature $T_c \gtrsim T_{dS}$.
The constituents may be then be composite. Starting from a small energy fluctuation, (\ref{EQN_mD}) sets an upper bound on the number of {\em pairings}, lowering $T_c$ by a factor $2^{-5/3}$ in doubling the mass of the constituents in each subsequent pairing \citep{van10}.

Ultra-light DM (\ref{EQN_mD}) is inherently dark by transparency, due to negligible cross-sections at the associated Compton wavelength to standard model interaction energies. 
Our bound (\ref{EQN_mD}) is consistent the estimate $m_Dc^2\gtrsim 10^{-22}$eV for wave-like dark matter ($\psi$DM) \citep{shi14,hui17,leu18,poz20,dem20,bro20,poz21,her19,poz23,bau22}, also known as fuzzy dark matter \citep{hu00,pee00,sik09}).
The condition $\lambda>r_t$ sets a smallest scale of dark matter structures consistent with the $\sim 10$kpc scale in recent ALMA surveys \citep{ino23}.

 At cosmic dawn, sensitivity of galaxy dynamics to $\sqrt{J}$ predicts early galaxy formation by rapid gravitational collapse, that may account for the JWST 'Impossible galaxies' \citep{van24b}. We identify this with reduced inertia across a $C^0$-transition across $a_{dS}$, below Newtonian inertia defined by the invariant mass-energy of a particle (\S6-7). 
Below $a_{dS}$, it accounts for radial accelerations in the bTFR with redshift dependent Milgrom parameter $a_0(z)$ (\ref{EQN_a0}). It represents the consistent IR coupling to position information (\ref{EQN_I}) as a primitive in general relativity (\ref{EQN_p1}).
By $\psi$CDM, however, first galaxies form {\em later} than CDM in $\Lambda$CDM \citep{shi14}, making the ultra high-redshift JWST galaxies even more `impossible'.

\subsection{Outlook}

High-resolution observations of $(H_0,q_0,S_8)$ promise
further observational tests of $J$CDM and the
scaling relations (\ref{EQN_scaling}) of $\Lambda$CDM parameters.
Confirmations beyond Table 1 promise to establish $J$CDM 
as the classical limit of a Big Bang quantum cosmology,
probed in late-time cosmology by the Local 
Distance Ladder. Specifically, we mention the following. 

\begin{itemize}
\item 
The reduced matter density $\Omega_{M,0}$ in (\ref{EQN_scaling}) is expected to alleviate $S_8$-tension, provided that $\sigma_8$ remains unchanged. This may be confirmed in a fit to the {\em Planck} power density of the CMB, now on the background (\ref{EQN_HJ}) of $J$CDM.
\item 
Accurate measurements of $q_0$ are essential to establish the order of the Hamiltonian energy constraint, i.e., second order in $J$CDM by (\ref{EQN_q0J}) versus first order in $\Lambda$CDM. These measurements are generally challenging due when derived from $H^\prime(z)$. Possibly, $q(z)$ may be determined from the BAO in galaxy surveys, e.g., by {\em Euclid}. 
\item 
The $6\sigma$ gap at the $C^0$-transition between Newtonian and anomalous galaxy dynamics (Fig. \ref{figRC}) signals reduced inertia across $a_{dS}$. In keeping with a three-flat Universe, such is to be preserved by the cosmological distribution of CDM, contributing to the gravitational field on scales {\em larger} but not smaller than $r_t$ (\ref{EQN_rt}). This condition points to ultra-light DM (\ref{EQN_mD}) and 
may be tested in high-resolution maps of the dark matter distribution \citep{ino23}.
\end{itemize}

While a theory of quantum cosmology continues to elude us, this outlook derived from two primitives $\Lambda_0$ and $I$ points to new physical properties the dark sector, 
presently dominated by a cosmic relic of de Sitter-Schouten heat
with no tension between early- and late-time cosmology \cite{van24d}.


\begin{acknowledgments}
We gratefully acknowledge the organizers of the workshop {\em Tensions in Cosmology} for creating a stimulating
interdisciplinary meeting during Corfu2023, and detailed comments on
the present manuscript from M.A. Abchouyeh.
\end{acknowledgments}

\bibliographystyle{apa}

\end{document}